# Stability of stratified two-phase flows in horizontal channels

I. Barmak[a)], A. Gelfgat, H. Vitoshkin, A. Ullmann, and N. Brauner

*School of Mechanical Engineering, Tel Aviv University, Tel Aviv 69978, Israel*

Linear stability of stratified two-phase flows in horizontal channels to arbitrary wavenumber disturbances is studied. The problem is reduced to Orr-Sommerfeld equations for the stream function disturbances, defined in each sublayer and coupled via boundary conditions that account also for possible interface deformation and capillary forces. Applying the Chebyshev collocation method, the equations and interface boundary conditions are reduced to the generalized eigenvalue problems solved by standard means of numerical linear algebra for the entire spectrum of eigenvalues and the associated eigenvectors. Some additional conclusions concerning the instability nature are derived from the most unstable perturbation patterns. The results are summarized in the form of stability maps showing the operational conditions at which a stratified-smooth flow pattern is stable. It is found that for gas-liquid and liquid-liquid systems the stratified flow with a smooth interface is stable only in confined zone of relatively low flow rates, which is in agreement with experiments, but is not predicted by long-wave analysis. Depending on the flow conditions, the critical perturbations can originate mainly at the interface (so-called "interfacial modes of instability") or in the bulk of one of the phases (i.e., "shear modes"). The present analysis revealed that there is no definite correlation between the type of instability and the perturbation wavelength.

## I. INTRODUCTION

Stratified flow is a basic flow pattern for gas-liquid and liquid-liquid two-phase flows in a gravitational field, where the lighter fluid flows above the heavier one. This flow pattern is frequently encountered in various important industrial processes. For example, gas-condensate pipelines operate primarily in the stratified flow regime. Clearly, the stratified two-phase flow regime can be achieved only for certain ranges of operating conditions for which this flow configuration is stable. Therefore, the knowledge of the stability limits is essential for the design and operation of pipelines and other process equipment that involve this flow pattern. Instability may result in transition from stratified-smooth to stratified-wavy flow. Interfacial waves in stratified flow were experimentally observed by Charles and Lilleleht (1965) and Yu and Sparrow (1969) and were studied in many following two-phase flow experimental works (see e.g., Andritsos and Hanratty, 1987; Tzotzi and Andritsos, 2013; Birvalski et al., 2014 and references therein). However, growing interfacial disturbances may lead to transition to other flow patterns (e.g., slug flow, annular flow). Therefore, stability analysis of stratified flow is considered a basic tool to be employed in the modelling of flow pattern transitions and for the prediction of the flow pattern map that is pertinent to the particular two-phase flow system of interest.

The exact formulation of transient flow in pipes is too complicated for conducting a rigorous stability analysis. An exact analytical solution for a steady two-phase laminar flow is available in the literature, where the velocity profiles are in the form of Fourier integrals in a bi-polar coordinate system (e.g., Ullmann et al., 2004,

---

[a)] Author to whom correspondence should be addressed. Electronic mail: ilyab@post.tau.ac.il.

Goldstein et al., 2015). However, a stability analysis requires a three-dimensional formulation and results in a too complicated problem. Therefore, a common approach is to use a transient one dimensional Two-Fluid (TF) mechanistic model for the stability analysis. Early studies on the stability of stratified gas-liquid flow in pipes assumed inviscid flow and considered only the transient momentum equation of the gas phase, ignoring the dynamics of the liquid phase (e.g., Kordyban and Ranov, 1970, Wallis and Dobson, 1973, Taitel and Dukler, 1976, Mishima and Ishii, 1980). Obviously, the prediction of the stratified flow boundary for a general two-phase system requires consideration of the inertia and viscous effects of both phases. A more rigorous approach that was followed in many later studies is to consider viscid flow and to use the 1D transient continuity and momentum equations of the two phases (e.g., Lin and Hanratty, 1986, Wu et al., 1987, Andritsos et al., 1989, Brauner and Moalem Maron, 1991,1993, Barnea and Taitel, 1994). However, the predictions obtained via the Two-Fluid model critically depend on the reliability of the closure relations used to model the base flow and the interaction of the base flow with the interfacial disturbance (e.g., steady and wave induced wall and interfacial shear stresses, velocity profile shape factors, see Kushnir et al., 2007, 2014). Moreover, the stability analysis is restricted to long wave perturbations, since long-wave perturbation is an inherent assumption when the Two-Fluid model is used.

An alternative approach is to conduct a rigorous stability analysis of a two-layer plane Poiseuille flow (i.e., Two Plates (TP) geometry), while considering all wave number perturbations in order to obtain insight into the mechanisms involved in the destabilization of stratified flow. Such a formulation allows for the Squire transformation (Hesla et al., 1986), so that only two-dimensional disturbances can be considered. In this approach, two types of instability are considered: shear-flow instability and interfacial instability. Shear-flow instability is due to the interaction of the flow with the channel walls (encountered also in single-phase Poiseuille flow, i.e., Tollmien-Schlichting waves), which leads to transition to turbulent flow in either of the phases for sufficiently large Reynolds number. This type of instability is commonly associated with short-wave instability. Interfacial instability is due to the interaction between the fluids at the interface, which results from energy transfer from the main flow to interfacial disturbances. In this case the instability is attributed to viscosity and/or density stratification (jump), and is commonly associated with (relatively) long-wave instability.

Boomkamp and Miesen (1996) offered a classification of ways of energy transfer from the base to the disturbed flow. According to that study, there are five different mechanisms by which energy transfer can be conducted. Each mechanism has its origins in one of the following flow features: density stratification, velocity profile curvature, viscosity stratification, shear effects, or a combination of viscosity stratification and shear effects. The stability analysis of stratified two-phase flow remains rather complicated even in the simpler two-plate geometry due to the multiplicity of the dimensionless parameters involved. Consequently, the studies conducted were usually limited to a narrow parameter range (e.g., Boomkamp and Miesen, 1996). A brief review on the history and most important results obtained for the considered problem is given below.

The linear stability of stratified plane horizontal Poiseuille flow of two superposed fluids with different viscosities was first studied theoretically in the classic work of Yih (1967), where long-wave two-dimensional analysis was carried out. It was shown that the viscosity stratification alone can cause interfacial instability for an arbitrary low Reynolds number, when the fluids occupy equal volumes in the channel. Yih introduced the concept of an interfacial mode, which can coincide with any streamline and is neutrally stable in single-fluid flow, but triggers instability when viscosity stratification exists.

The interface may be unstable also to short-wavelength perturbations as was shown by Hooper and Boyd (1983). For the case of equal densities and zero surface tension, they obtained exact dispersion relation for disturbances of arbitrary wavelength. Taking an unbounded Couette flow of two fluids of different viscosity as an example, it was shown additionally that presence of rigid boundaries does not play an essential role in the flow instability (in contrast to the case of plane Poiseuille flow of a homogeneous fluid). It was also demonstrated that the effect of surface tension is always stabilizing, whereas a density difference may have either stabilizing or destabilizing effect.

The numerical study of Yiantsios and Higgins (1988) accounted for density stratification, different fluid layers thickness, the effects of interfacial tension and gravity. They showed that a shear perturbation mode, which is essentially a short-wavelength disturbance of the Tollmien-Schlichting type, modified by the interfacial effects, can destabilize stratified flow at sufficiently large Reynolds numbers. For the single fluid case, the shear mode originated at the channel walls is the only source of instabilities, but two-phase flow may be unstable to more than one mode. Tilley, Davis and Bankoff (1994) extended the study of Yiantsios and Higgins (1988). They studied the influence of the channel height and the (heavy) layer thickness on the flow stability in horizontal and inclined channels. Their analysis of the dominant unstable modes demonstrates the qualitative differences between single- and two-layer systems. The effect of inclination was further studied by Vempati et al., (2010), but they paid a limited attention to horizontal flows. The most recent paper on the topic of linear instability in two-phase horizontal flow was published by Kaffel and Riaz (2015). They focused on spectral characteristics and eigenfunction patterns (amplitude of the stream function perturbation) related to shear and interfacial modes of instability and the interaction between them in the unstable region.

The above studies contribute to the understanding of the nature of instability and the main mechanisms that are responsible for the flow destabilization. However, the analyses conducted in those studies were not used to obtain physical interpretations that may lead to improvement of the modelling of flow pattern transitions.

In this study, the linear stability of stratified two-phase flows in horizontal channels with respect to arbitrary wavenumber disturbances is explored. Temporal disturbances are considered, so that the mathematical problem reduces to solution of a series of generalized eigenvalue problems. For the sake of further physical interpretations and practical applications, this rigorous analysis tool is applied for predicting the stability boundaries on the flow pattern maps of particular gas-liquid and liquid-liquid systems, thus extending previous results of long-wave analysis by Kushnir et al., (2014). The critical disturbances that are responsible for triggering the instability (i.e., the most unstable disturbance mode) are identified. For characterization of the instability mode associated with the critical disturbances two-dimensional contours of the stream function and the amplitude of the velocity perturbations are examined. The obtained results offer a better physical insight into the phenomenon of two-phase flow instability and may help to improve the modelling of flow pattern transitions, which are usually based on the simplified 1D transient two-fluid models.

## II. PROBLEM FORMULATION

We consider a stratified two-layer flow of two immiscible incompressible fluids in a horizontal channel (i.e., zero angle of inclination from the horizontal, $\beta = 0°$). The flow is assumed to be isothermal and is driven by an imposed pressure gradient. The flow configuration is sketched in Figure 1. The interface between the fluids, labeled as $j = 1,2$ (1 – lower phase, 2 – upper phase), is assumed to be flat in the undisturbed base flow state. Under this assumption, the model allows for a plane-parallel solution, in which position of the interface is an unknown value (to be determined below).

The flow in each liquid is described by the continuity and momentum equations that are rendered dimensionless (see Kushnir et al., 2014), choosing for the scales of length and velocity the height of the upper layer $h_2$ and the interfacial velocity $u_i$, respectively. The time and the pressure are scaled by $h_2/u_i$, and $\rho_2 u_i^2$, respectively. The dimensionless continuity and momentum equations are

$$div\,\mathbf{u}_j = 0,$$
$$\frac{\partial \mathbf{u}_j}{\partial t} + (\mathbf{u}_j \cdot \nabla)\mathbf{u}_j = -\frac{\rho_1}{r\rho_j}\nabla p_j + \frac{1}{\text{Re}_2}\frac{\rho_1}{r\rho_j}\frac{m\mu_j}{\mu_1}\Delta \mathbf{u}_j - \frac{1}{\text{Fr}_2}\mathbf{e}_y, \qquad (1)$$

where $\mathbf{u}_j = (u_j, v_j)$ and $p_j$ are the velocity and pressure of the fluid $j$, $\rho_j$ and $\mu_j$ are the corresponding density and dynamic viscosity; $\mathbf{e}_y$ is the unit vector in the direction of the y-axis (the direction of gravity).

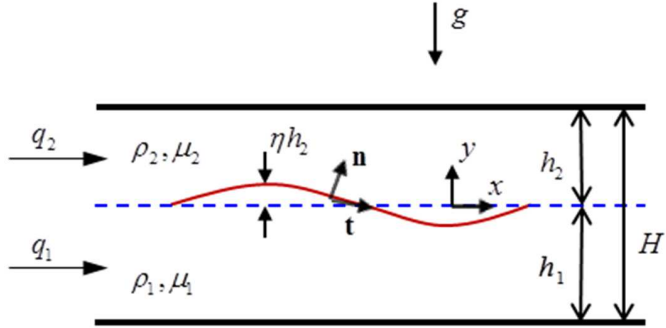

FIG. 1. Configuration of stratified two-layer flow in a horizontal channel.

In the dimensionless formulation the lower and upper phases occupy the regions $-n \leq y \leq 0$, and $0 \leq y \leq 1$, respectively, where $n = h_1/h_2$. Other dimensionless parameters: $\text{Re}_2 = \rho_2 u_i h_2/\mu_2$ is the Reynolds number, $\text{Fr}_2 = u_i^2/gh_2$ is the Froude number, $r = \rho_1/\rho_2$ and $m = \mu_1/\mu_2$ are the density and viscosity ratios. Since the horizontal two-phase flow allows for the Squire transformation (Hesla et al., 1986), we assume that most unstable disturbances are two-dimensional and should therefore be considered in the stability analysis.

The velocities satisfy the no-slip boundary conditions at the channel walls

$$\mathbf{u}_1(y = -n) = 0, \quad \mathbf{u}_2(y = 1) = 0. \qquad (2)$$

Boundary conditions at the interface $y = \eta(x,t)$ require continuity of velocity components and the tangential stresses, and jump of the normal stress due to the surface tension (the square brackets denote the jump of expression value across the interface)

$$\mathbf{u}_1(y=0) = \mathbf{u}_2(y=0), \qquad (3)$$

$$[\mathbf{t \cdot T \cdot n}] = \left[ \frac{m\mu}{\mu_1} \left\{ \left( \frac{\partial u}{\partial y} + \frac{\partial v}{\partial x} \right) \left( 1 - \left( \frac{\partial \eta}{\partial x} \right)^2 \right) - 4 \frac{\partial u}{\partial x} \frac{\partial \eta}{\partial x} \right\} \right] = 0, \qquad (4)$$

$$[\mathbf{n \cdot T \cdot n}] = \left[ p + \frac{m\mu}{\mu_1} \frac{2\mathrm{Re}_2^{-1}}{1+\left(\frac{\partial \eta}{\partial x}\right)^2} \left( \frac{\partial u}{\partial x}\left(1-\left(\frac{\partial \eta}{\partial x}\right)^2\right) + \left( \frac{\partial u}{\partial y} + \frac{\partial v}{\partial x} \right)\frac{\partial \eta}{\partial x} \right) \right]$$

$$= \mathrm{We}_2^{-1} \frac{\frac{\partial^2 \eta}{\partial x^2}}{\left(1+\left(\frac{\partial \eta}{\partial x}\right)^2\right)^{3/2}}, \qquad (5)$$

where $\mathbf{n}$ is the unit normal vector pointing from lower into upper phase, $\mathbf{t}$ is the unit vector tangent to the interface, $\mathbf{T}$ is the stress tensor. An additional dimensionless parameter $We_2 = \rho_2 h_2 u_i^2 / \sigma$ is the Weber number, and $\sigma$ is the surface tension coefficient.

Additionally, the interface displacement and the normal velocity components at the interface satisfy the kinematic boundary condition

$$v_j = \frac{D\eta}{Dt} = \frac{\partial \eta}{\partial t} + u_j \frac{\partial \eta}{\partial x}. \qquad (6)$$

**III. THE BASE FLOW**

The base flow is assumed to be steady, laminar, and fully developed. Assuming that the horizontal velocity $U(y)$ varies only with the vertical coordinate, $y$, the exact steady-state solution for the (dimensionless) velocity profiles reads (e.g., Ullmann et al., 2003, Kushnir et al., 2014):

$$U_1 = 1 + a_1 y + b_1 y^2 \text{ for } -n \leq y \leq 0, \quad U_2 = 1 + a_2 y + b_2 y^2 \text{ for } 0 \leq y \leq 1, \qquad (7)$$

where

$$a_1 = \frac{a_2}{m}; \quad a_2 = \frac{m-n^2}{n^2+n}; \quad b_1 = -\frac{m+n}{(n^2+n)m}; \quad b_2 = -\frac{m+n}{n^2+n}.$$

To apply collocation method based on the Chebyshev polynomials (defined in the interval $[0,1]$), we introduce a new coordinate $y_1 = (y+n)/n$, whereby $0 \leq y_1 \leq 1$ for the part of the channel occupied by the lower phase, and $y_2 = y$, $0 \leq y_2 \leq 1$ for the upper phase (which remains unchanged). After substitution of the new coordinate the velocity profile in the lower phase reads

$$U_1 = 1 + a_1 y + b_1 y^2 = 1 + a_1 (y_1 - 1) n + b_1 (y_1 - 1)^2 n^2$$
$$= \underbrace{1 - a_1 n + b_1 n^2}_{\tilde{c}_1} + \underbrace{(a_1 n - 2b_1 n^2)}_{\tilde{a}_1} y_1 + \underbrace{b_1 n^2}_{\tilde{b}_1} y_1^2. \quad (8)$$

It is convenient to use the lower (heavy) phase holdup, $h = h_1 / H$, instead of thickness ratio $n = h_1 / h_2$. The heavy phase holdup can be found by solving the following algebraic equation, $F(q, m, h) = 0$ (e.g., Ullmann et al., 2003a):

$$\frac{mq(1-h)^2 \left[ (1+2h)m + (1-m)h(4-h) \right] - h^2 \left[ (3-2h)m + (1-m)h^2 \right]}{4h^3 (1-h)^3 \left[ h + m(1-h) \right]} = 0. \quad (9)$$

Equation (9) can be represented as a 4-th order algebraic equation in $h$, using the Martinelli parameter

$$X^2 = \frac{(-dP/dx)_{1S}}{(-dP/dx)_{2S}} = m \cdot q$$

$$(X^2 + 1)(m-1)h^4 + 2X^2 (3 - 2m) h^3 + \left[ 3X^2 (2m-3) - 3m \right] h^2$$
$$+ 4X^2 (1-m) h + X^2 m = 0. \quad (10)$$

Thus, the primary flow solution for horizontal flow is fully determined by two dimensionless parameters, the viscosity ratio $m = \mu_1 / \mu_2$ and the flow rate ratio $q = q_1 / q_2$. Here $q_j$ is the feed flow rate of phase $j$ and $(-dP/dx)_{jS} = 12 \mu_j q_j / H^3$ is the corresponding superficial pressure drop of phase $j$ (i.e., when flowing alone in the channel of height $H = h_1 + h_2$). Another important characteristic of the flow – the dimensionless pressure drop can be calculated as

$$\tilde{P} = \frac{dP/dx}{(-dP/dx)_{2S}} = \frac{3mq(1-h)^2 - 4mh(1-h) - h^2}{4h(1-h)^2 \left[ (1+2h)m + (1-m)h(4-h) - 3h \right]}. \quad (11)$$

The interfacial velocity can also be found

$$\tilde{u}_i = \frac{u_i}{U_{2S}} = \frac{6h(1-h)(-\tilde{P})}{m(1-h) + h}, \quad (12)$$

where $U_{2S} = q_2 / H$ is the superficial velocity of the upper phase.

Henceforth, we use the superficial velocity and the flow rate concepts interchangeable due to consideration of channels of constant height.

## IV. LINEAR STABILITY ANALYSIS

In the following we study linear stability of above plane-parallel solution with respect to infinitesimal, two-dimensional disturbances. The perturbed velocities and pressure fields are written as $u_j = U_j + \tilde{u}_j$, $v_j = \tilde{v}_j$, $p_j = P_j + \tilde{p}_j$, and $\eta = \tilde{\eta}$ for the dimensionless disturbance of the interface. The disturbed velocities are conveniently represented by the corresponding stream function $\left( \tilde{u}_j = \partial \psi_j / \partial y \,;\, \tilde{v}_j = -\partial \psi_j / \partial x \right)$, and an exponential dependence of the perturbation in time is assumed

$$\begin{pmatrix} \psi_j \\ \tilde{p}_j \\ \eta \end{pmatrix} = \begin{pmatrix} \phi_j(y) \\ f_j(y) \\ H_\eta \end{pmatrix} e^{(ikx+\lambda t)} \,;\, \begin{pmatrix} \tilde{u}_j \\ \tilde{v}_j \end{pmatrix} = \begin{pmatrix} \phi'_j \\ -ik\phi_j \end{pmatrix} e^{(ikx+\lambda t)}, \qquad (13)$$

where $\phi_j$, $f_j$ and $H_\eta$ are the perturbation amplitudes, $k$ is the dimensionless real wavenumber ($k = 2\pi h_2 / l_{wave}$, with $l_{wave}$ being the wavelength) and $\lambda$ is the complex time increment. After substitution and linearization of the original equations and boundary conditions (1)-(6), the problem is reduced to the Orr-Sommerfeld equations, written here in the eigenvalue problem form

$$0 \leq y_1 \leq 1: \qquad \lambda D_1 \phi_1 = \left[ ik\left(-U_1 D_1 + \frac{U_1''}{n^2}\right) + \frac{1}{\mathrm{Re}_1} D_1^2 \right] \phi_1, \qquad (14)$$

$$0 \leq y_2 \leq 1: \qquad \lambda D \phi_2 = \left[ ik\left(-U_2 D + U_2''\right) + \frac{1}{\mathrm{Re}_2} D^2 \right] \phi_2, \qquad (15)$$

where

$$D\phi = \phi'' - k^2 \phi; \quad D^2 \phi = \phi^{IV} - 2k^2 \phi'' + k^4 \phi;$$
$$D_1 \phi = \frac{\phi''}{n^2} - k^2 \phi; \quad D_1^2 \phi = \frac{\phi^{IV}}{n^4} - 2k^2 \frac{\phi''}{n^2} + k^4 \phi. \qquad (16)$$

The linearized boundary conditions are obtained by means of Taylor expansions of $\eta$ around its unperturbed zero value (see Segal, 2008, for more details)

$$y_1 = 1,$$
$$y_2 = 0: \qquad \lambda H_\eta = -ik\left(\phi_2 + U_2 H_\eta\right),$$
$$\text{where} \quad H_\eta = \frac{\phi'_2(0) - \phi'_1(1)/n}{U'_1(1)/n - U'_2(0)} \qquad (17)$$

$$y_1 = 1, \quad \lambda\left(r \cdot \frac{\phi_1'(1)}{n} - \phi_2'(0)\right) = ik\left[-\left(k^2 \operatorname{We}_2^{-1} + \frac{(r-1)}{\operatorname{Fr}_2}\right) \cdot H_\eta\right.$$
$$y_2 = 0: \quad + r\left(-U_1\frac{\phi_1'(1)}{n} + \frac{U_1'(1)}{n}\phi_1(1)\right) + \left(U_2\phi_2'(0) - U_2'(0)\phi_2(0)\right)\right] \quad (18)$$
$$+ \frac{1}{\operatorname{Re}_2}\left[m\left(\frac{\phi_1'''(1)}{n^3} - 3k^2\frac{\phi_1'(1)}{n}\right) - \left(\phi_2'''(0) - 3k^2\phi_2'(0)\right)\right]$$

$$y_1 = 0 \atop (y = -n) \qquad \phi_1 = \phi_1' = 0 \qquad (19)$$

$$y_2 = 1 \qquad \phi_2 = \phi_2' = 0 \qquad (20)$$

$$y_1 = 1, \atop y_2 = 0: \qquad \phi_1(1) = \phi_2(0) \qquad (21)$$

$$y_1 = 1, \atop y_2 = 0: \qquad m\left[\frac{\phi_1''(1)}{n^2} + k^2\phi_1(1) + \frac{U_1''}{n^2}H_\eta\right] = \phi_2''(0) + k^2\phi_2(0) + U_2''H_\eta \qquad (22)$$

Note that the third terms on both sides of equation (22) are equal in case of horizontal flow and cancel out.

The temporal linear stability is studied by solving the system of differential system (14), (15) and (17)-(22) assuming an arbitrary wavenumber for each given set of the other parameters. Defining the time increment as a complex eigenvalue $\lambda = \lambda_R + i\lambda_I$, we note that $\lambda_R$ determines the growth rate of perturbation. The flow is unstable when at least one of the $\lambda_R$'s is positive. Neutral stability corresponds to $\max(\lambda_R) = 0$. The flow is considered to be stable, when real parts of all the eigenvalues are negative. The phase speed of the perturbation is $-\lambda_I/k$, where $\lambda_I$ is the wave angular frequency.

**V. NUMERICAL METHOD**

Following several previous studies, the Chebyshev collocation method is used to discretize the Orr-Sommerfeld equations and the boundary conditions. This numerical method uses Chebyshev polynomials to approximate the solution in each sublayer as a truncated Chebyshev series

$$\phi^{(j)} = \sum_{i=1}^{N} d_i^{(j)} T_i(y), \qquad (23)$$

where $j=1,2$ – phase index; $d_i^{(j)}$ -unknown coefficients, $T_i(y)$ - shifted Chebyshev polynomials of the $i$-th degree over the interval $0 \le y \le 1$. The equations are evaluated at the roots of the Chebyshev polynomial of N-th order (collocation points), reducing the differential problem to the generalized eigenvalue problem

$$\lambda \cdot A \cdot d = B \cdot d, \qquad (24)$$

where the matrices order is of *2N+1*. The eigenvalue problem is solved by the QR algorithm (Francis, 1962). Accurate numerical results were compared with the long-wave asymptotic solution of Kushnir et al. (2014) (see Table I). The numerical convergence of the method was verified separately and is illustrated in Table II.

Most of the results below are computed using the truncation number $N=50$, that according to Table II yields 6 correct decimal digits of the critical superficial velocity.

TABLE I. Comparison of the present numerical results and asymptotic solution of Kushnir et al. (2014) for the critical superficial velocities. Calculation with N=50.

|  | Horizontal channel $H=0.02\,\text{m}$ | $U_{2S}$, m/s | $U_{1S}$, m/s | |
|---|---|---|---|---|
|  |  |  | Asymptotic | Numerical |
| Air-water flow | $r=1000$ $m=55$ | 5 | 0.01633291 | 0.0163241 |
|  |  | 2.5 | 0.03104910 | 0.03104204 |
|  |  | 0.25 | 0.05228360 | 0.05227238 |
|  |  | 0.15 | 0.88270124 | 0.88266123 |
| Liquid-Liquid flow | $r=1.25$ $m=0.5$ | 0.001 | 0.14987568 | 0.14985647 |
|  |  | 0.05 | 0.28990455 | 0.28984558 |
|  |  | 0.3 | 0.28623871 | 0.2862288 |
|  |  | 1 | 0.73131243 | 0.73135426 |

TABLE II. Convergence of the critical superficial velocity of the heavy phase $U_{1s}$ with the increase of truncation number N.

| Order of Chebyshev polynomials, *N* | Horizontal air-water flow: $H=0.02\,\text{m},\ r=1000,\ m=55,\ \sigma=0.072\,\text{N/m}$ | |
|---|---|---|
|  | $U_{2S}=0.25\,\text{m/s},\ k=10^{-5}$ | $U_{2S}=2\,\text{m/s},\ k=3$ |
| 25 | 1.7051426 | 0.010672387 |
| 50 | 1.7051135 | 0.010611218 |
| 75 | 1.7051134 | 0.010611217 |
| 100 | 1.7051134 | 0.010611217 |

## VI. RESULTS AND DISCUSSION

All wavelength linear stability analysis of horizontal liquid-liquid and gas-liquid flows was carried out to study stability limits of stratified flows and to reveal the destabilization mechanisms involved. Although the base flow characteristics are determined by only two dimensionless parameters ($m, q$), five parameters, i.e., $m, q, r, \text{Re}_{2S}$ or $\text{Fr}_{2S}$, and $\text{We}_{2s}$ govern the linear stability of the base flow and its further non-linear evolution. Here $\text{Re}_{js} = \rho_j U_{js} H / \mu_j$ and $\text{Fr}_{js} = U_{js}^2 / (gH)$ are the superficial Reynolds and Froude numbers respectively, and $\text{We}_{js} = \rho_j U_{js}^2 H / \sigma$ is the Weber number. The large number of parameters makes the overall parametric analysis practically unfeasible. Therefore, in the present work we set the values of the physical properties at several representative examples and study the stability of the flow by varying flow rates in each fluid layer. In all cases the upper layer is considered to be lighter than the lower layer, such that Rayleigh-Taylor instability is not encountered. Along with presenting the all wavelength stability boundaries, we are interested in comparison with the analytical results obtained for the long-wave limit. The most unstable perturbations described by the leading eigenfunctions are also reported and their patterns are discussed, which allows arriving at additional conclusions on the nature of instability.

There are several mechanisms, which can be responsible for destabilization of the flat interface solution. One of them is a shear flow instability, which is encountered also in single-phase Poiseuille flow and leads to transition to turbulent flow for Reynolds numbers higher than critical. This instability originates near the channel walls and is related to amplification of short wavelength Tollmien-Schlichting waves in either of the phases. The presence of this instability mode already proves that the long-wave analysis is insufficient for determining the stability boundaries of stratified two-phase flow.

In two-phase flow it may occur that instability originates at the interface, and is associated with viscosity and/or density stratification (e.g., Yih, 1967; Kushnir et al., 2014). Such instability is viewed as a result of interaction of the flows in the two layers, which are connected through the velocity and viscous stresses boundary conditions at the interface (Tilley, 1994). As a result, the flow can become unstable for lower flow rates (lower superficial Reynolds number) than in plane Poiseuille flow. Viscosity stratification $(m \neq 1)$ produces a discontinuity (jump) across the interface in the primary flow velocity gradient $U_j'$, which leads to energy transfer from the primary flow to the disturbed flow and causes the 'viscosity induced' instability. According to Hooper and Boyd (1983) and Boomkamp and Miesen (1996), this is the dominant mechanism for the so-called 'interfacial instability'. It should be emphasized that since the stability problem is linear, several independent destabilizing mechanisms may be observed, which will appear as distinct unstable eigenfunctions.

**A. Liquid-liquid systems**

First, we consider two liquid-liquid systems with different viscosity ratios and with identical density ratios and channel heights. The stability map for horizontal flow with a more viscous $(m = 0.5)$ light phase (oil) is shown in Figure 2. The surface tension is neglected. The stability map is plotted for dimensional superficial velocities, for the channel of 2 cm height, the oil density and dynamic viscosity are $\rho_2 = 800 \text{kg/m}^3$ and $\mu_2 = 0.0005 \text{Pa} \cdot \text{s}$, respectively. In the framework of long the long-wave analysis (black dashed line in Figure 2) this flow is shown to be stable for any upper (light and more viscous) phase flow rate, provided the flow rate of the heavy (less viscous) phase is sufficiently low (for details see Kushnir et al, 2014). According to the long-wave analysis, at low oil superficial velocities there exists a critical heavy phase superficial velocity that depends on density ratio between phases and flow rate of light phase (black dashed line in Figure 2). In the region of large flow rates of both phases, when viscosity effects surpass the effect of gravity, the long-wave stability boundary approaches the zero-gravity stability boundary (see Kushnir et al, 2014). This behavior of the long-wave stability boundary is common for all horizontal two-layer systems, and observed below also for gas-liquid flows. Under zero gravity conditions, the long-wave instability is characterized by a critical holdup, which is determined by the viscosity ratio, $h_{cr} = \sqrt{m}/(1+\sqrt{m})$ (Yiantsios and Higgins, 1988), and corresponds to a critical flow rate ratio of $q_{cr} = (\sqrt{m}+m)/(1+\sqrt{m})$. At the critical flow rate ratio, the velocity gradient of the base flow is continuous across the interface, whereby interfacial shear stress is zero and the average velocities of the two layers are equal. Under zero-gravity conditions (i.e., either $g=0$ or $r=1$) the flow is stable with respect to long-wave disturbances in the region where the flow in the layer of the more viscous phase is faster. Namely, for $m<1$ the stable region corresponds to $q<q_{cr}, 0<h<h_{cr}$ ( i.e., below the green dashed line in Figure 2), while for $m>1$ the stable region is above this line $(q>q_{cr}, 1>h>h_{cr})$. Accordingly, the stabilizing effect of gravity on long-wave perturbations are observed in Figure 2 in the region of the faster less viscous (heavier) layer (i.e., above the $q_{cr}$ line), as this region would be unstable under zero gravity conditions. For an additional verification of the numerical results we calculated the stability boundary for $k \to 0$ (blue squares), which successfully compare with the analytical results for the long-wave instability (black dashed line).

Considering perturbations of all possible wavelengths, we obtain different results. The flow is stable only in a bounded region of the phases flow rates (within the red curve in Figure 2), which is much smaller than that obtained by the long-wave approach. This observation means that intermediate and short wave disturbances are amplified stronger for lower flow rates than the long waves. More details can be seen in Figures 3 and 4. Figure 3 presents stability diagrams for zero and non-zero surface tension including also the critical wavenumbers (normalized by the channel height, $k_H = 2\pi H/l_{wave}$) for both cases. Figure 4 presents two examples of the growth rate dependence on the wavenumber (at points A and B in Figure 3).

Comparing instabilities at low $U_{2S}$ for finite and infinite Weber numbers, we see that at zero surface tension $(We \to \infty)$ the critical mode wavenumber is considerably larger than unity, indicating short-wave instability. Including a realistic surface tension in the model (finite $We$) we expect a stabilization effect, since it becomes more difficult to perturb the interface. Obviously, the surface tension suppresses short wave perturbations, while leaves the long waves almost unaffected. A stabilization effect of the surface tension is indeed observed (Figure 3), however, its effect on the critical oil and water superficial velocities is rather moderate. The main effect of the surface tension is on the wavenumber of the critical perturbation. As surface tension stabilizes the short waves, the critical perturbation is shifted to lower wavenumber, i.e., longer waves are responsible for triggering instability. This is demonstrated in Figure 4, where the growth rate of the perturbations vs. the wavenumber at points A and B are compared. As demonstrated in Figure 4(b), a slight increase of the superficial velocity of one of the phases beyond the critical conditions (i.e., in the unstable region), results in positive values of the growth rate of perturbations in a range of wavenumbers around the critical one, indicating flow instability.

Single phase limits for laminar flow of light phase and laminar flow of heavy phase are plotted in Figure 3. When oil flow rate tends to zero, the neutral stability curve approaches the single phase limit for water flow determined by the critical Reynolds number ($Re_{Cr} = 5772$, critical wavenumber $k_H = 1.02$ (e.g., Orszag, 1971)). The wavenumber of the most unstable mode also tends to that of single phase flow. Therefore, it can be concluded that this is a shear mode of instability, which is associated with the short wavelength Tollmien-Schlichting wave, modified by the presence of the interface. The introduction of a thin layer of oil (small holdup) to the (single phase) water flow has a stabilizing effect. On the other hand, if a thin heavy layer of water is added to the (single phase) oil flow, a destabilizing effect is observed.

The stability map for horizontal oil-water flow with a less viscous $(m=2)$ light phase (oil: $\rho_2 = 800\,\text{kg/m}^3$, $\mu_2 = 0.0005\,\text{Pa}\cdot\text{s}$, $\mu_1 = 0.001\,\text{Pa}\cdot\text{s}$) is shown in Figure 5. In accordance with the discussion above (with reference to Figure 2), when long wave perturbations are considered $(k \to 0)$, the region where the water layer is faster would be stable under zero gravity conditions (i.e., above the $q_{cr}$ line). In this case, the stabilizing effects of gravity are observed in the region where the oil (less viscous) layer is faster. Accordingly, the flow is stable in an unbounded region of large holdups of water (low oil flow rates), and is limited only by the flow rates of the less viscous oil phase. However, owing to the short wavelength perturbations, the flow is stable only in a bounded range of the water (more viscous phase) flow rates, and also in a reduced range the oil flow rates. The neutral stability curve in this case looks similar to what was observed in the flow with inverse viscosity ratio $m=0.5$. The stabilizing effect of the surface tension (expansion of stable region) is significant only in the region of similar flow rates (holdups) of the two phases. It can be concluded that in oil-water flows the effect of surface tension on the stable region is rather moderate.

An additional physical insight into the flow destabilization mechanisms can be drawn from the patterns of the most unstable perturbations (eigenfunctions). The absolute values of the latter are shown as amplitudes of perturbations of the stream function and its derivative ($abs(\phi)$ and $abs(\phi')$, red and blue solid lines, respectively) in Figure 7. The vertical and horizontal velocity disturbances are proportional to the disturbances of the stream function and its derivative (see Eq. (13)), and $abs(\phi(0))$ is related to the interface displacement amplitude (see Eq. (17)). The base flow velocity profiles are also shown (green dashed lines), as in single phase flow the maximum of $abs(\phi)$ coincides with the location of the maximal axial velocity (i.e., the channel center). All the profiles are normalized by their values at the interface. Note that $abs(\phi')$ is discontinuous across the interface and is normalized by $abs(\phi'(0))$ in the light phase. The plots are presented for the five characteristic points (A, B, C, D, and E) marked in Figure 6, corresponding to high, intermediate and low holdups of the heavier liquid, respectively.

Assuming that the maximum of the perturbation amplitude corresponds to the location where instability evolves, we can make some additional speculations on the flow destabilization. Thus, for high (low) holdups of heavy phase (points A and E, respectively), instability arises in the bulk of the light (heavy) phase layer (see Figures 7(a) and (i)), and can be associated with a shear mode. This can be interpreted as an effect of large shear at the walls, while the effect of viscosity stratification at the interface (located close to the channel wall) is balanced by gravity stabilization. Nevertheless, the presence of the interface may result in a shift of the maximum disturbance from the location of the maximal velocity profile towards the interface (e.g., point A, Figure 7(a)), or in a local additional maximum near the interface (e.g., point E, Figure 7(i)). Moreover, the two-phase flow becomes unstable for lower superficial velocities than the oil single-phase flow. In addition, the eigenfunction derivative undergoes a rapid change not only near the walls, as in the single-phase case, but also near the interface (Figures 7(b) and (j)). In contrast, for intermediate holdups (point C) the maximum of the perturbation amplitude is at the interface (Figure 7(e)). While this may imply that the critical perturbation in this case corresponds to an interfacial mode, the variation of the eigenfunction derivative resembles that of a shear mode (i.e., maxima near the channels wall, and a minimum at the location of the maximal velocity profile, see Figure 7(f)). However, point C actually represents a particular case, where the maximal velocity of the primary flow is near the interface, whereby the velocity gradient is almost continuous across the interface. On the other hand, at points B and D the maximal disturbance is at the interface (Figure 7(c) and (g)), and the eigenfunction derivative attains a maximal value at (or close to) the interface (Figure 7(d) and (h)). This suggests that in these two points the instability can be characterized as an interfacial mode. Nevertheless, all of the critical disturbances for this case study are intermediate or short waves ($k_H$ is of the order of 1, see Figure 5).

While Figure 7 shows the amplitude of the stream function disturbance, no information can be obtained about its phase. Further insight can be obtained by examining the two-dimensional contours of the stream function of the critical disturbance. These contours represent the real part of the perturbation (Eq.(13)) at a particular time (e.g., $t=0$), and reads

$$\mathrm{Re}(\psi_j) = \mathrm{Re}(\phi_j(y)e^{ikx}) = \mathrm{Re}(\phi_j(y))\cdot\cos(kx) - \mathrm{Im}(\phi_j(y))\cdot\sin(kx). \qquad (25)$$

Figure 8 shows the 2D contours of the stream function perturbation (in the range $0 \leq x \leq 3l_{wave}$) obtained for the conditions corresponding to points A-E in Figures 6,7). The perturbations are normalized by the absolute value of their amplitude at the interface. Basically, the perturbation is a series of identical (in terms of shape) vortices of $l_{wave}/2$ width along the flow direction. Their core corresponds to the maximum value of the stream function. In the case of a thin layer of oil ( i.e. point A, Figure 8(a)) the vortex core is observed in the bulk of the water (lower) layer (corresponding to the maximum in Figure 7(a)), providing an additional evidence in favor of shear mode. In contrast to single phase flow, here the vortex is tilted in the flow direction (the phase of the disturbance is dependent on the location in the flow). The streamlines (black curves in Figure 8) have abrupt bends at the interface (due to discontinuity in the stream function derivative, see Figure 7(b)). For lower holdup (point B, Figure 8(b)), an additional core (local maximum) is formed, and some of the streamlines have an "8" shape, preserving a single vortex structure. The classification of this perturbation is rather ambiguous. Yet, in view of Figures 7(c) and (d), the interface plays a leading role in the flow destabilization, therefore this case can be classified as an interfacial mode. The critical perturbation for the flow with an intermediate holdup (point C, Figure 8(c)) is represented by vertical vortices perpendicular to the flow direction (location-independent perturbation phase), thus further substantiating the similarity of this particular case with single-phase flow instability (as discussed with reference to Figure 7(e),(f)). A further decrease in the holdup results in vortices tilting against the flow (Figure 8(d)), where the vortex core is still at the interface (indicating an interfacial mode). However, for thin water layers, the vortex core is located in the bulk of the oil layer (see Figure 8(e)) and can be classified as a shear mode. The similarity in the pattern of the critical perturbations observed in the cases of low and high holdups seems reasonable in view of the small differences in the fluids properties (i.e., the density and viscosity ratios are close to 1). It should be noted that further studies on the mode evolution and interaction of unstable modes beyond the neutrally stable boundary, where the flow becomes (linearly) unstable with respect to several modes, cannot be performed in the framework of the linear stability analysis and, therefore, is out of scope of this work.

The stability boundary predicted with two-fluid (TF) model of Kushnir et al. (2007) is compared with present analysis in Figures 9 (a) and (b). Note that long-wave perturbation is the underlying assumption when the TF model is used for stability analysis. It was shown by Kushnir et al. (2007) that the exact long-wave boundary can be obtained via TF model when several modifications are introduced in this model. These include introducing (destabilizing) terms due to wave-induced tangential (wall and interfacial) shear stresses in phase with the wave slope, and shape-factor in the inertia terms to account for the velocity profiles in the two layers. However, as discussed above, the exact long-wave stability boundary (dashed black curve in Figure 9) does not predict a bounded stable region of the flow as reported in experimental flow pattern maps. At the same time discarding those modifications in TF model (i.e., ignoring those wave induced shear stresses and assuming plug flow in the two layers with shape factors of 1) a confined stable region of stratified flow is predicted (orange dotted curve). Although it is the simplest model, its predictions appear to be closer to the results of the present analysis that considers the stability of all wavenumbers for reconstructing the stability boundary. Thus, our results can confirm the usefulness of the simple TF model for obtaining an estimation of the stability limits of stratified two-phase flows.

### B. Gas-liquid systems

Gas liquid systems are characterized by high viscosity and density ratios, and the stability limits of the stratified gas-liquid flows have been extensively studied in the literature. Considering air-water flow as an example, the heavy phase (water) is significantly more viscous $(m=55)$ than the light phase (air), and the density ratio $(r=1000)$ is much higher than in liquid-liquid systems. Consequently, as shown in Figure 10, the stabilizing effect of gravity is stronger and the long-wave stable region is extended to larger light phase flow rates. Stability analysis for an arbitrary wavenumber of the perturbation yields a confined stable region, which also appears to be wider than that of liquid-liquid flows. The modes with short and intermediate wavelengths are the most unstable and responsible for the instability onset almost in the entire range of holdups. Exceptions are flows with similar superficial velocities of the phases (holdup around 0.5) or very thin water layers, where the exact neutral stability curve coincides with the long-wave stability boundary (Figure 11). Note that a jump in the critical wavenumber along the stability boundary reflects the fact that there may be more than one dangerous mode that can become unstable, and the wavenumber indicated on the stability map is the most unstable mode. The stabilization effect of the surface tension is observed to be more significant than in the liquid-liquid systems considered above (but still rather moderate). This can be attributed not only to the higher surface tension of the air-water system, but also to the fact that the critical perturbations correspond to shorter waves that are dumped stronger by surface tension. Here too, the main effect of surface tension is the shift of the critical perturbation to smaller wavenumbers.

The observed flow pattern transitions from stratified flow with a smooth interface to other regimes (elongated bubble / slug, pseudo-slugs, or stratified wavy flow) in channels of a similar size are presented in Figure 12. Instability of thick water layers (i.e., high water flow rates and relatively low gas flow rates) results in transition to elongated-bubble or slug flow (e.g., Barnea et al., 1980). The results clearly indicate that this transition is not associated with long-wave instability. For this size of a channel $(H = 0.02\,\text{m})$, the long waves can trigger instability for very thin water layers, where transition from stratified-smooth to stratified-wavy was observed. The long waves are found to be the critical disturbance also in the region of intermediate liquid holdups where transition from stratified-smooth to pseudo-slugs was observed, implying that this transition is associated with long-wave instability.

The loss of the flow stability may be initiated in the bulk of one or both phases (shear mode), and may thus be associated with laminar/turbulent transition in the corresponding phases, or at the interface between the fluids (interfacial mode). When compared with the stability limits of air and water single phase flow (corresponding to $\text{Re}_{Cr} = 5772$, critical wavenumber $k_H = 1.02$, Figure 11), one can observe that the water flow is stabilized by the presence of a thin (less viscous and lighter) gas layer at the upper wall. For the other extreme case of a thin water layer, the two-phase stability boundary almost coincides with the stability boundary of single phase air flow, implying that instability is driven by growing perturbations in the bulk of the air layer. However, due to the presence of interface and surface tension, this process is characterized by longer wavelength critical disturbance.

To study the locations at which instability evolves, the five points along the neutral stability curve are taken (see Figure 12). The amplitudes of perturbations of the stream function and its derivative for the point A (very thin layer of the air above the water) are shown in Figures 13(a) and (b). The maxima in the amplitude of both the vertical and horizontal velocity disturbances (perturbations of the stream function and its derivative, respectively) are at the interface, whereas in the original base flow the maximum velocity is located within the water layer. In this case the thin layer of air flow above the water has a stabilizing effect on the shear mode instability due to the diminishing shear at the upper boundary. Upon increasing the water flow rate, instability is triggered at the water-air interface, suggesting that the flow is destabilized by a short wave interfacial mode.

For higher flow rates of air (point B) the most unstable mode is a short wave with the maximum disturbance amplitude in the air layer, close both to the interface and to the location of the maximal base flow velocity (see Figures 13(c) and (d)). A further decrease in the holdup (increase in the air flow rate) brings us to the region where very long waves are the most unstable perturbations (point C). The disturbance in this case can be characterized as interfacial mode, however another smaller maximum in the bulk of air is observed (see Figures 13(e) and (f)). In the region of small water flow rates (point D, Figures 13(g) and (h)) the disturbance maximum within the air layer becomes dominant. This occurs since a sufficiently thin layer of the more viscous phase (water) plays a role of wall for the air flow, and only a local (smaller) maximum is preserved at the interface. The most unstable wavelength in this case is still small $(k_H = 2.9)$, but there is the second dangerous long wave mode, which subsequently becomes the most unstable for lower water holdups (point E, Figures 13(i) and (j)). Yet, in this case instability evolves in the bulk of the air layer, implying shear mode instability, however, with a long wave. Inspection of Figure 13 indicates that due to the large viscosity ratio, the disturbance in the axial velocity attains a maximum value at the interface (except in the particular case where the maximum velocity of the primary flow is located at the interface and interfacial shear is zero). The general conclusion that can be drawn based on these results is that the value of the critical wavenumber cannot be used to classify the instability as a shear mode or an interfacial mode. Such a characterization can be made only based on the examination of the disturbance stream function amplitude.

The effect of downscaling and upscaling of the channel size on the flow stability is examined by studying the results obtained for air-water flow in channels of 0.002m and 0.2m height compared to those obtained in the 0.02m channel. The stability maps are presented in Figure 14(a) and (b). As seen from all three cases studied, the critical superficial water velocity for low gas flow rates is not that sensitive to channel height and is associated with short-wave instability. In contrast, for low liquid flow rates, the neutral stability curve is dependent on channel's height. When the channel size is increased from 0.02m to 0.2m, the critical disturbances correspond to short waves even for thin water layers. In this case the critical gas superficial velocity was found to correspond to a critical superficial Reynolds number $\left(\text{Re}_{2S}^{Cr} \approx 7800\right)$, whereby increasing the channel size results in a decrease of the critical air superficial velocity $(U_{2S} \propto 1/H)$. On the other hand, upon reducing the channel size form 0.02m to 0.002m, a major part of the stability boundary is associated with long-wave instability. In this case (see Kushnir et al., 2014) the critical air superficial velocity is found to correspond to a constant gas Froude number $(\text{Fr}_{2S}^{Cr} = U_{2S} \cdot \left[(r-1)gH\right]^{-0.5} \approx 0.35)$, whereby the critical air superficial velocity decreases with reducing the channel size $(U_{2S} \propto H^{0.5})$. These critical Fr number and Re number are insensitive to the value of the surface tension.

The widest stable region is obtained for $H = H_{Cr} = 0.025$m channel, when the critical air velocity corresponds both to the above $\text{Fr}_{2S}^{Cr}$ and $\text{Re}_{2S}^{Cr}$, and the very long waves $(k \to 0)$ are still the critical perturbation for thin water layers. In this case the flow becomes unstable at a critical air velocity of $U_{2S}^{Cr} = 5.7$ m/s (for $U_{1S} = 10^{-4}$ m/s, $h = 0.05$). This result seems to agree with the data on air-liquid flows in lab scale systems, where the air velocity for transition from stratified-smooth to stratified-wavy was found to be about 5 m/s at atmospheric pressure (e.g., Andritsos and Hanratty, 1987).

The above results can be presented in a form of general expressions for the critical channel height and the corresponding maximum critical gas superficial velocity:

$$H_{Cr} \approx \left[ \frac{\left(\text{Re}_{2S}^{Cr}\right)^2 \mu_2^2}{\left(\text{Fr}_{2S}^{Cr}\right)^2 \rho_2^2 g(r-1)} \right]^{1/3}, \qquad (26)$$

$$U_{2S}^{Cr} = \frac{\text{Re}_{2S}^{Cr} \mu_2}{\rho_2 H_{Cr}} = \left[(r-1)gH_{Cr}\right]^{1/2} \text{Fr}_{2S}^{Cr}. \qquad (27)$$

For channels larger than $H_{Cr}$, the effect of the channel size on the transitional $U_{2S}$ should be evaluated based on $\left(\text{Re}_{2S}^{Cr} \approx 7800\right)$, whereas for $H < H_{Cr}$ the scaling should be based on $\text{Fr}_{2S}^{Cr}$. Moreover, the above scaling rules apply also for predicting the effect of pressure on the critical gas velocity. Applying Eqs. (26)-(27) for high-pressure systems $(r < 1000)$, indicates that for each particular density ratio there is a critical channel height for which the stable stratified flow extends over a largest range of air flow rates. The scaling based on the Fr number (i.e., $U_{2S} \approx (\rho_{2o}/\rho_2)^{0.5} (U_{2S})_o$ suggested by Andritsos and Hanratty, 1987, subscript $o$ denotes atmospheric conditions) may be applicable only to channels of $H < H_{Cr}$. For $H > H_{Cr}$, the scaling based on Re suggests that $U_{2S} \propto 1/\rho_2$. At elevated pressures, the critical channel size and the corresponding maximal $U_{2S}^{Cr}$ decrease according to Eqs. (26)-(27), implying that the range of gas flow rates for which stratified flow with a smooth interface can be maintained in field operations, which are associated with large channel sizes and elevated pressures, is much smaller than that obtained in lab scale experiments. Similarly, the range of stable (smooth) stratified flow when downscaling to micro channels significantly diminishes, which is in a general agreement with experimental findings (e.g., Kawahara et al., 2002, Qu et al., 2004, Ide et al., 2007).Note that the critical Fr and Re are however dependent on the viscosity ratio and should be identified for the particular gas-liquid system under consideration.

Another interesting result concerns the values of the critical superficial water velocity obtained at low gas superficial velocities. With increasing the channel size from 0.02m to 0.2m, the critical water superficial velocity somewhat increases, whereas for single phase water flow the critical water flow rate would decrease. This results in an extended region of water flow rates where the added thin gas layer stabilizes the shear mode instability in the water. In fact, in larger channels, the critical shear in the gas layer (for a specified (low) gas superficial velocity) is reached at a higher holdup, hence at higher water superficial velocity. For the same reason, in mini/micro channels, the critical water flow rate is already smaller than that obtained in single phase water flow. The stability of the flow in region of small air superficial velocities is surface tension dominated and as shown below, resembles that obtained in micro gravity conditions.

The limiting case of a system under zero gravity condition is of particular interest. It corresponds either to absence of gravity or to flow of fluids with equal densities. As seen in Figure 15(a), for a long wave perturbation, the flow remains stable for holdups of water larger than the critical ($h_{cr}$), where the water (the more viscous phase) is faster (i.e., above the $q_{cr}$ line). However, consideration of arbitrary perturbations changes this picture. Without surface tension (Figure 15(b)), the neutral stability curve of any nonzero k is located in the region below the $q_{cr}$ line (where the air (less viscous phase) is faster and the long waves are unstable). The larger is the wavenumber the smaller is the stable region. For sufficiently large wavenumbers no stable region exists, which is in agreement with asymptotic solution of Yiantsios and Higgins (1988). Hence, when all wavenumbers are considered, smooth stratified flow cannot be stabilized without surface tension. With surface tension, a rather small stable region is obtained in the part of the long-wave stable region corresponding to sufficiently low air and water flow rates (Figure 16). Similarly, in case the heavy phase is the less viscous one and surface tension is included (e.g., $m = 0.5$, Figure 17), the flow can be stable under zero gravity conditions in part of the long-wave stable region, which in this case corresponds to $q < q_{cr}$ and $h < h_{cr}$ (the region where the light more viscous phase is faster).

**VII. CONCLUSIONS**

The linear stability of a steady plane-parallel two-layer flow was studied for liquid-liquid and gas-liquid systems. The analysis was carried out taking into consideration all possible wavenumbers and without any additional simplifications. The results are presented as stability maps accompanied by the wavenumbers and spatial profiles of the most unstable perturbations.

According to the present study, horizontal gas-liquid and liquid-liquid stratified flows can be stable to arbitrary wavenumber perturbations only in a confined zone of relatively small flow rates. This is in agreement with experimental observations, but not predicted by long-wave analysis (see Kushnir et al., 2014). Nevertheless, identification of systems and conditions where a long-wave disturbance is the critical one for triggering instability is of importance, as under such conditions the long-wave analytical solution (or the Two-Fluid model) can be conveniently applied for predicting the stable stratified flow boundaries.

 Short wave instability is a characteristic of thin layers of the less viscous fluid, whereas intermediate and long-wave perturbations may be dominant in triggering instability of small holdups of the more viscous fluid. Depending on the flow conditions, the critical perturbation can evolve mainly at the interface (so-called "interfacial mode" instability") or in the bulk of one of the phases (i.e., "shear mode"). However, it was shown that there is no definite correlation between the type of instability and the perturbation wavelength. In particular, long waves do not necessarily imply interfacial mode instability. A classification to shear or interfacial mode can be made only based on examination of the pattern of the disturbance stream function.

Additionally, the effect of the channel height on the stability of gas-liquid flow was studied. It was revealed that in small channels and small holdups, a long wave is the critical disturbance. In this case the channel size effect on the critical gas superficial velocity for the onset of instability is scaled by a critical gas (superficial) Froude number. On the other hand, in large channels, short/intermediate waves are the critical disturbances, and the channel size effect on the critical gas velocity is scaled by a critical gas (superficial) Reynolds number. Long waves become the critical perturbations in channel heights less than a critical one. The latter can be determined for each particular density ratio (i.e., operational pressure). An increase of the channel height above the critical one results in a decrease of the critical gas flow rate, owing to short wave instability. In channel of the critical height, a stable stratified flow region can be obtained for the largest range of gas flow rates corresponding to both the critical Froude and the critical Reynolds numbers. The values of the critical Fr and Re numbers are however dependent on the viscosity ratio. These findings are of practical importance especially for upscaling lab-scale (and low pressure) data on the stratified-smooth boundary to field operational conditions.

The surface tension was shown to affect more the short waves instability. However, due to the existence of other growing modes with larger wavelength, the surface tension effect on the stability boundaries is found to be rather small. The only exceptions are system under zero-gravity condition and flows in microchannels, where capillarity becomes significant. In the absence of a gravitational force, the surface tension plays a leading role in the flow stabilization, and for sufficiently small flow rates the flow is stable to all disturbances.

Yih, C. S., "Instability due to viscosity stratification," J. Fluid Mech. 27, 337–352 (1967).

Yu, H. S., Sparrow, E. M., "Experiments on two-component stratified flow in a horizontal duct," J. Heat Transfer 91, 51-58 (1969).

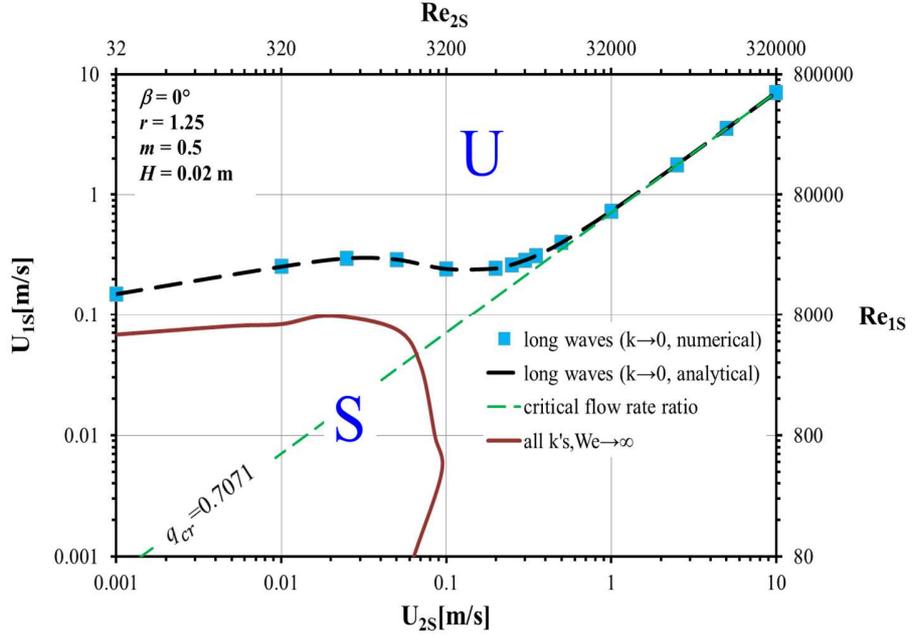

FIG. 2. Stability boundary for horizontal liquid-liquid flow without surface tension (We→∞), the light phase is more viscous $\left(m = 0.5;\ \rho_2 = 800\,\text{kg/m}^3;\ \mu_2 = 5\cdot 10^{-4}\,\text{Pa}\cdot\text{s}\right)$.

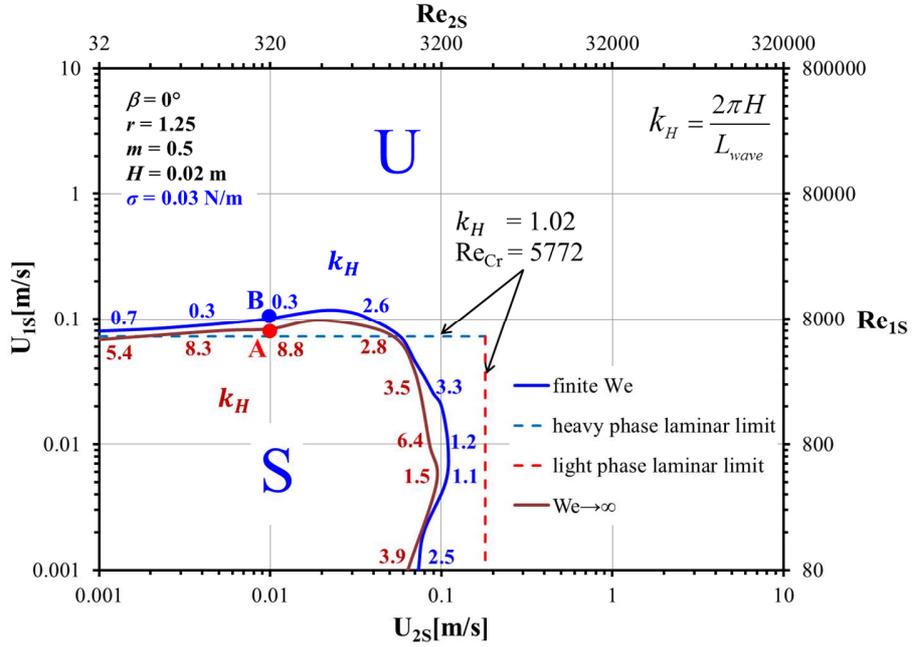

FIG. 3. Stability boundaries for horizontal liquid-liquid flow $\left(m = 0.5;\ \rho_2 = 800\,\text{kg/m}^3;\ \mu_2 = 5\cdot 10^{-4}\,\text{Pa}\cdot\text{s}\right)$ with (finite We) and without surface tension (We→∞).

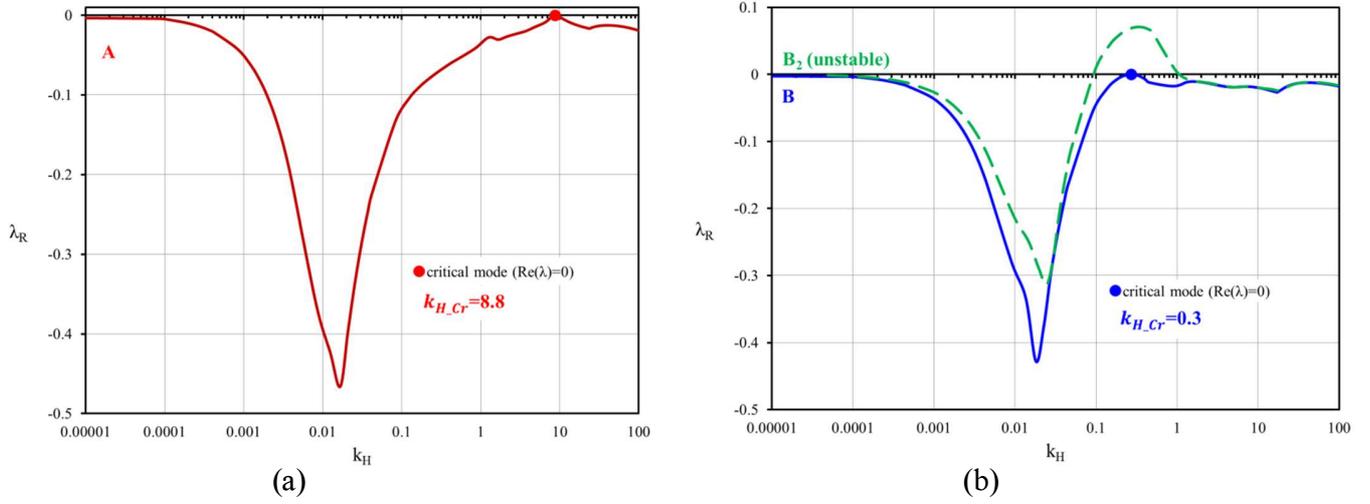

FIG. 4. Growth rate of perturbation vs. wavenumber at points on the neutral stability curves. (a) At point A $(U_{1s} = 0.083\,\text{m/s}; U_{2s} = 0.01\,\text{m/s}; h = 0.76)$ for the flow without surface tension; (b) at point B $(U_{1s} = 0.101\,\text{m/s}; U_{2s} = 0.01\,\text{m/s}; h = 0.779)$ for the flow with surface tension (Figure 3). Growth rates of perturbations at point B2 (flow with surface tension: $U_{1s} = 0.12\,\text{m/s}; U_{2s} = 0.01\,\text{m/s}; h = 0.79$, which is in the unstable region in the vicinity of point B) are illustrated by green dashed line.

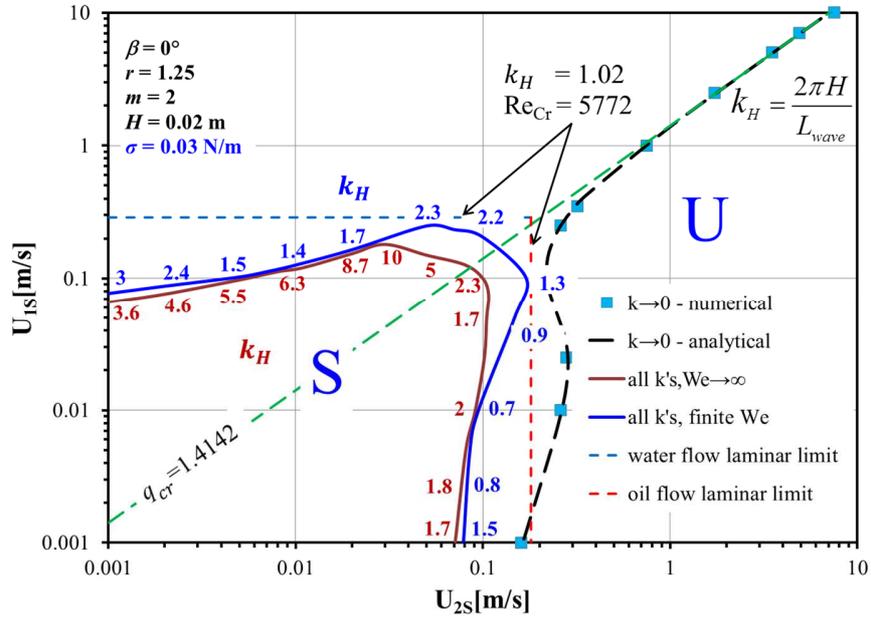

FIG. 5. Stability map for horizontal oil-water flow $(m = 2; \rho_2 = 800\,\text{kg/m}^3; \mu_2 = 5\cdot 10^{-4}\,\text{Pa}\cdot\text{s})$ with (finite We) and without surface tension (We→∞).

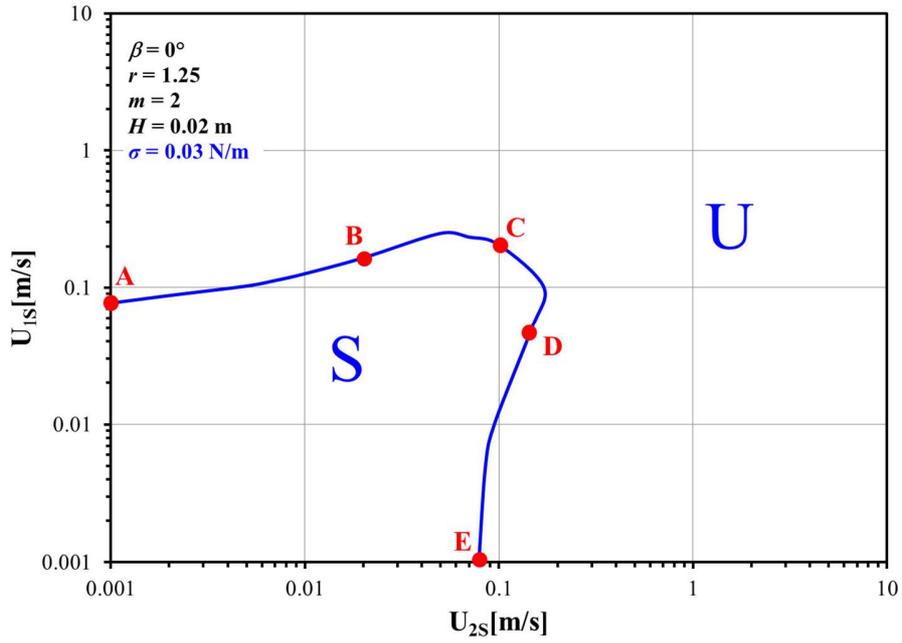

FIG. 6. Sampling points for studying the most dangerous perturbation in the horizontal oil-water flow

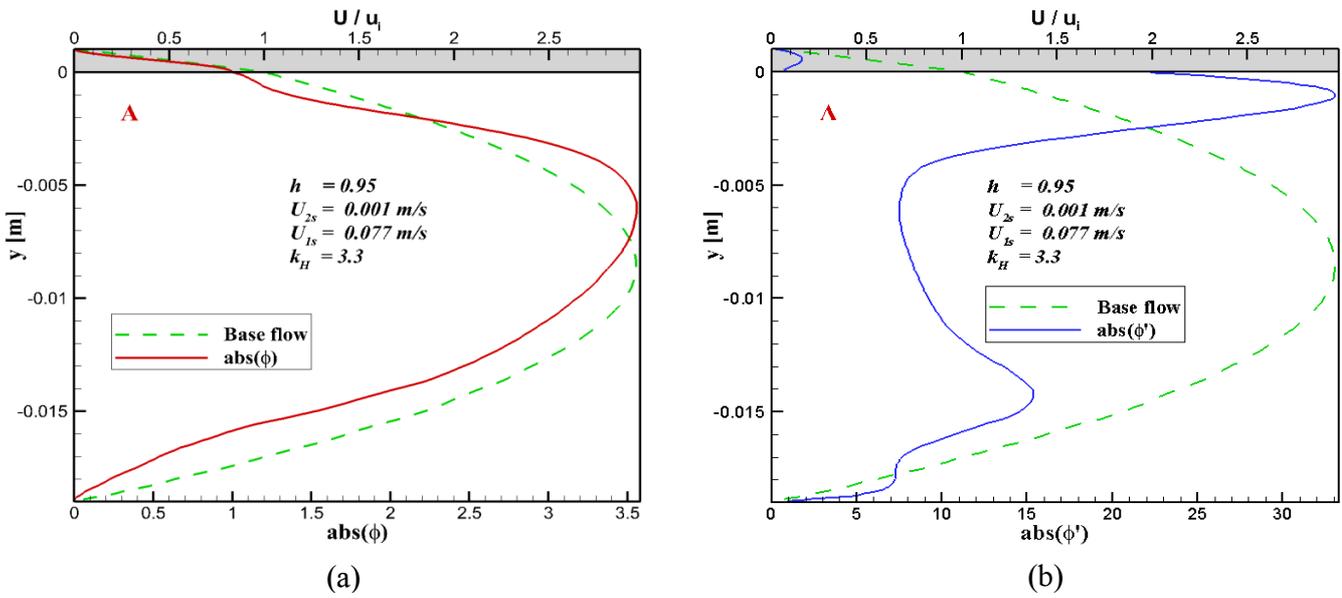

(a)            (b)

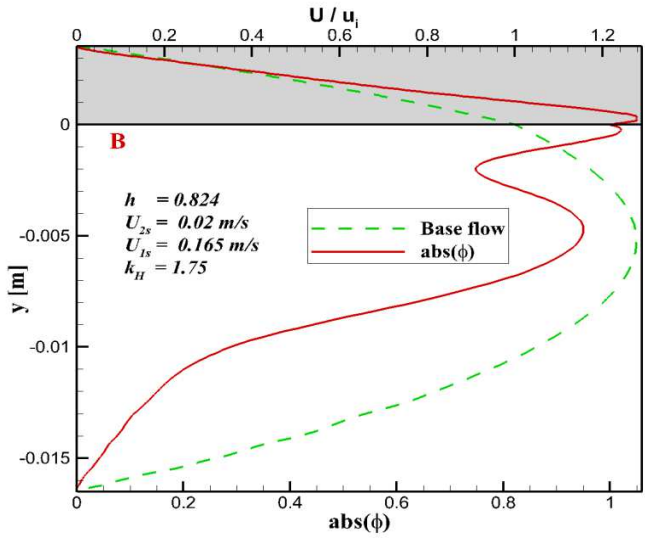
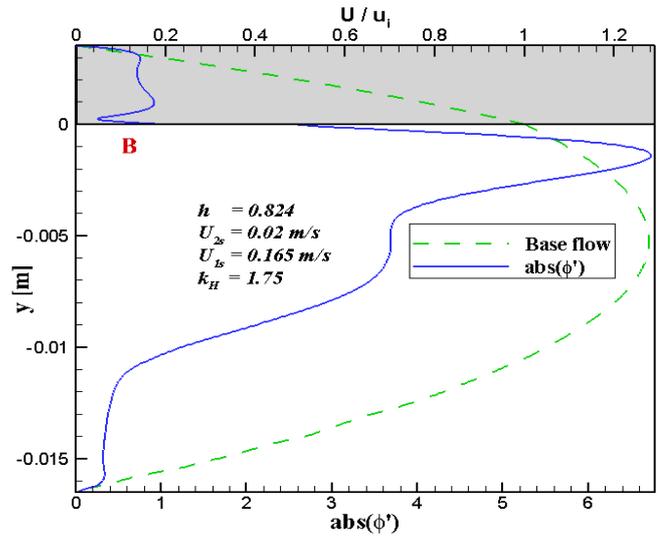

(c)                          (d)

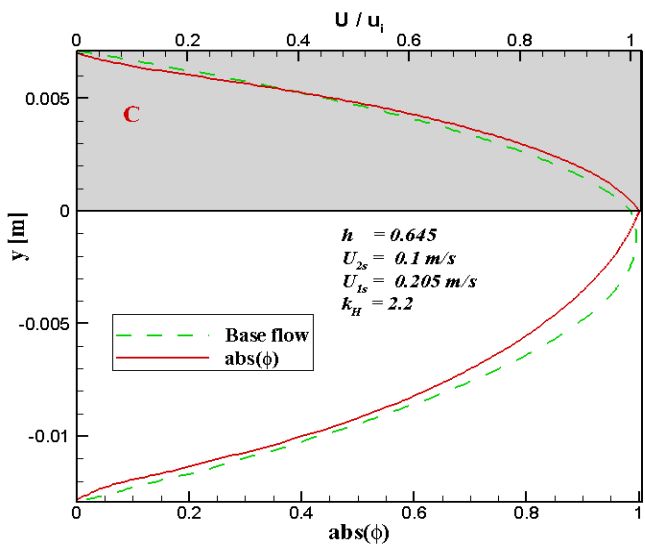
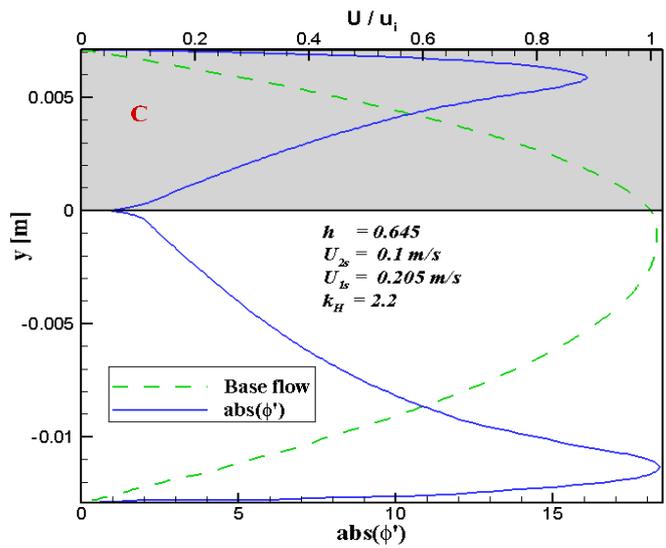

(e)                          (f)

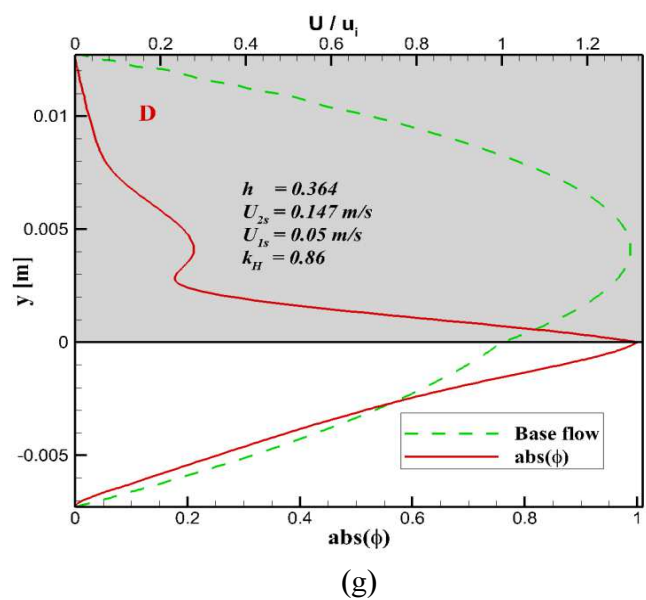
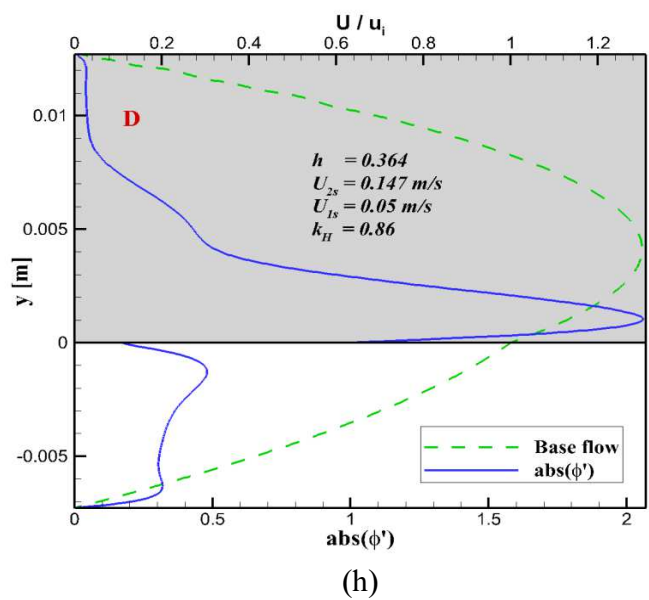

(g)             (h)

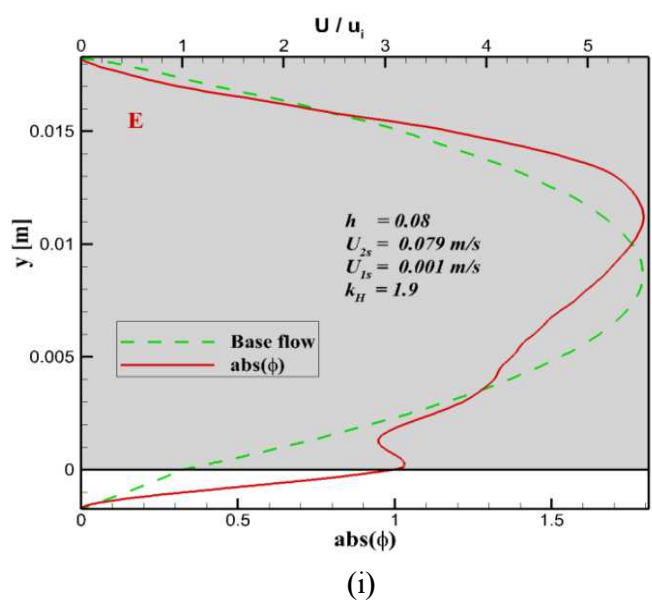
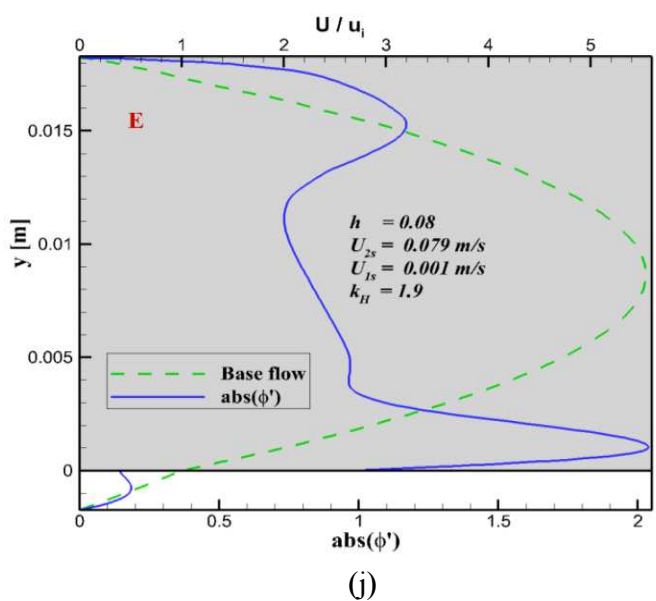

(i)             (j)

FIG. 7. Amplitudes of the critical perturbations of the stream function ((a), (c), (e), (g), and (i): eigenfunction, red solid lines) and its derivative ((b), (d), (f), (h), and (j): blue solid lines), and base flow velocity profile (green dashed lines) at points A-E (see Fig. 6); $y < 0$ – water; $y > 0$ (shaded region) – oil.

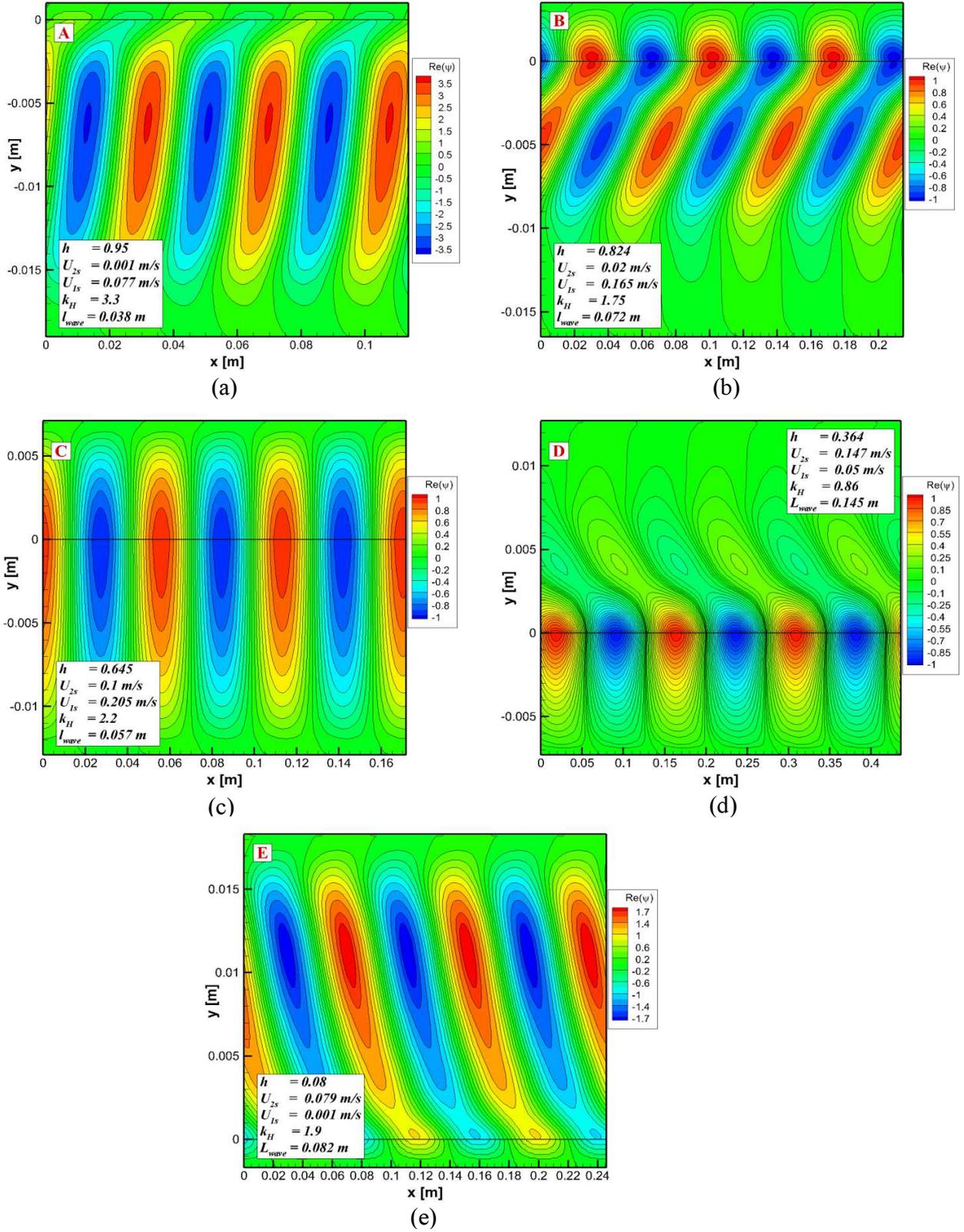

FIG. 8. Contours of the stream function perturbations ($\text{Re}(\psi)$, (a) – (e)) at points A-E (see Fig. 6); y < 0 – water; y > 0 – oil; $0 \leq x \leq 3 l_{wave}$.

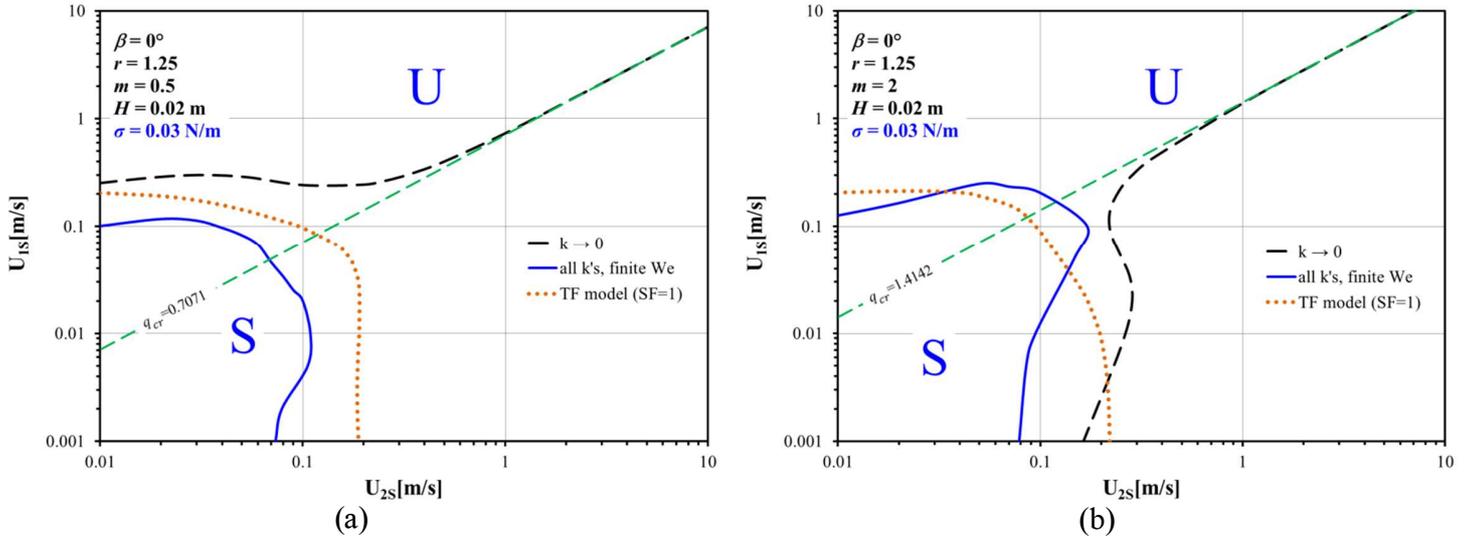

FIG. 9. Comparison of the exact analysis and two-fluid (TF) model of Kushnir et al. (2007) (assuming plug flow in the two layers with shape factors (SF) of 1) stability maps for liquid-liquid flows: (a) – light phase (oil) is more viscous $(m = 0.5)$; (b) – heavy phase (water) is more viscous $(m = 2)$.

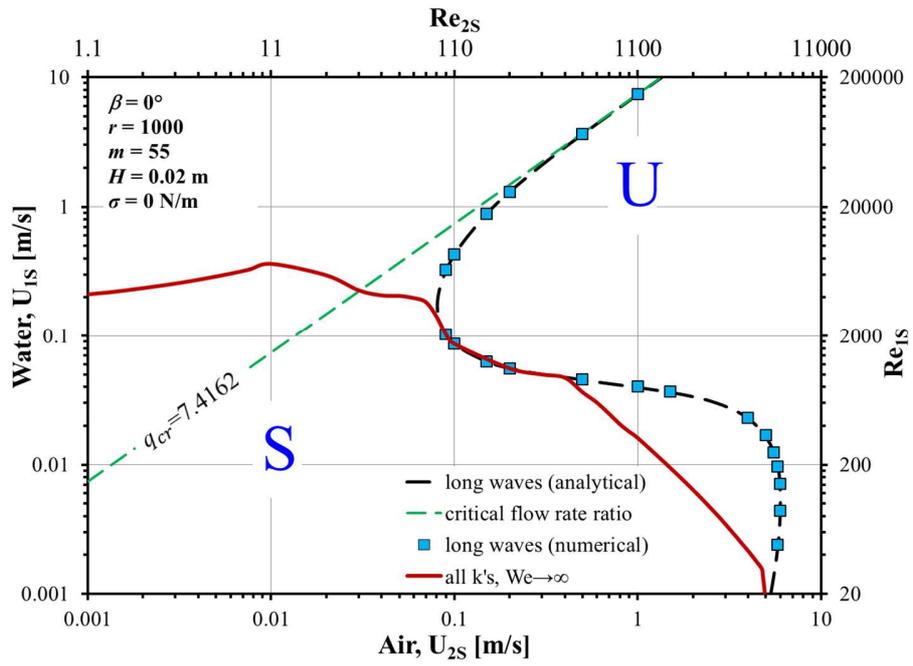

FIG. 10. Stability map for horizontal air-water system without surface tension (We→∞).

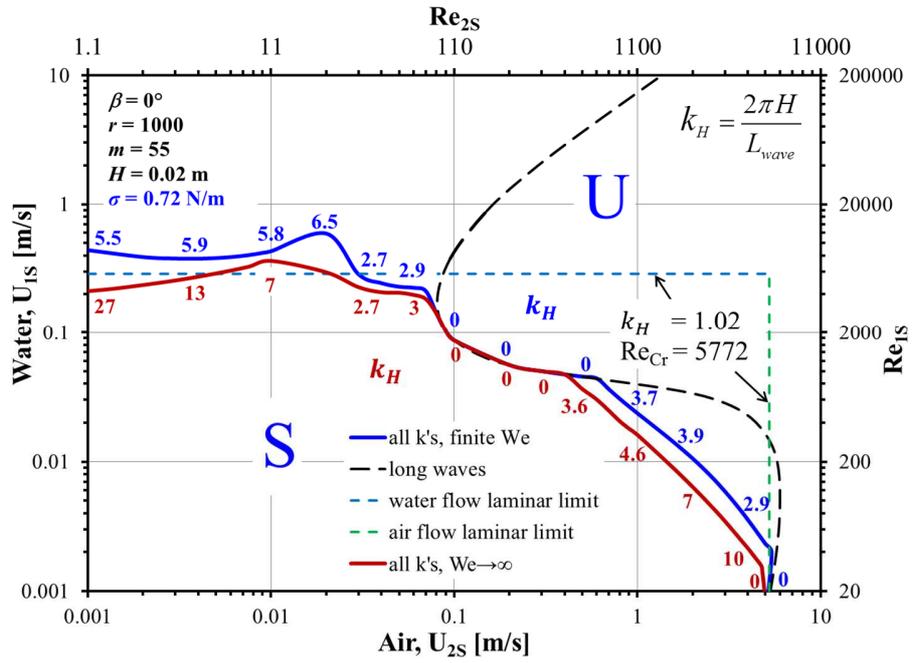

FIG. 11. Stability boundaries for horizontal air-water flow and air and water single phase laminar limits.

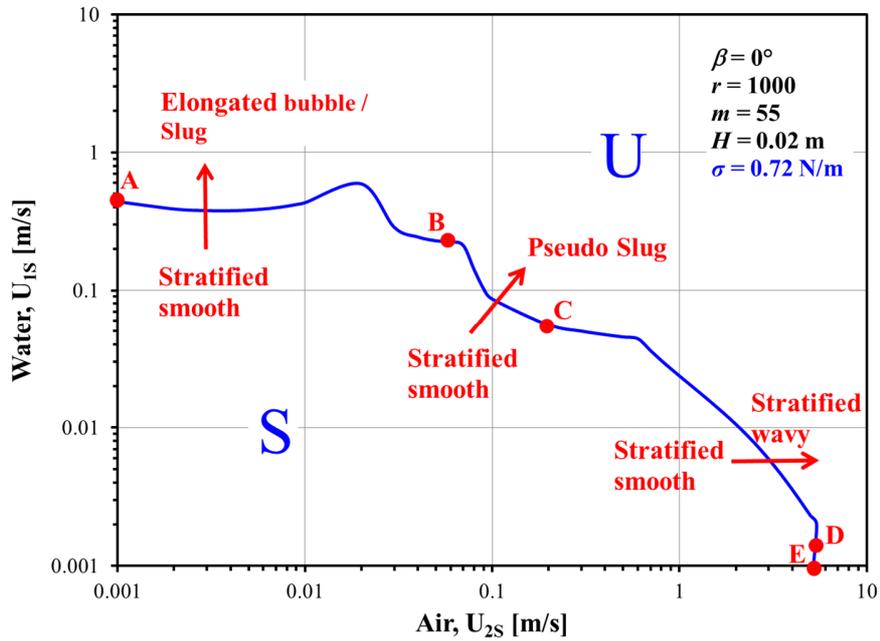

FIG. 12. Flow transition across stability boundary and sampling points for studying the critical perturbations in the horizontal air-water flow.

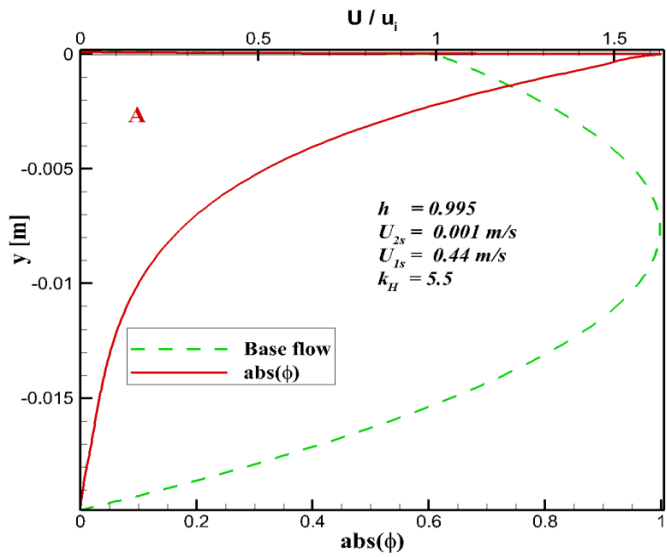
(a)

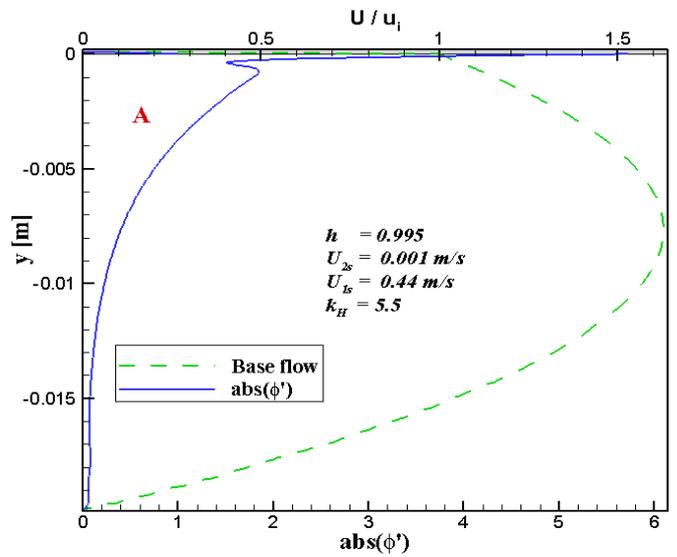
(b)

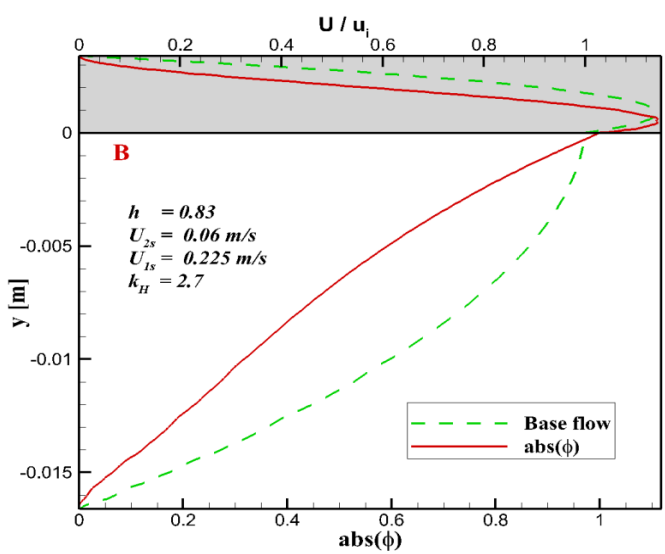
(c)

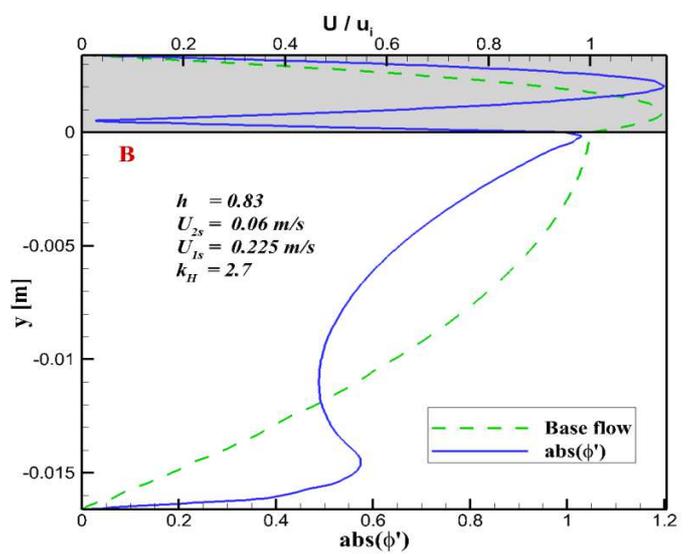
(d)

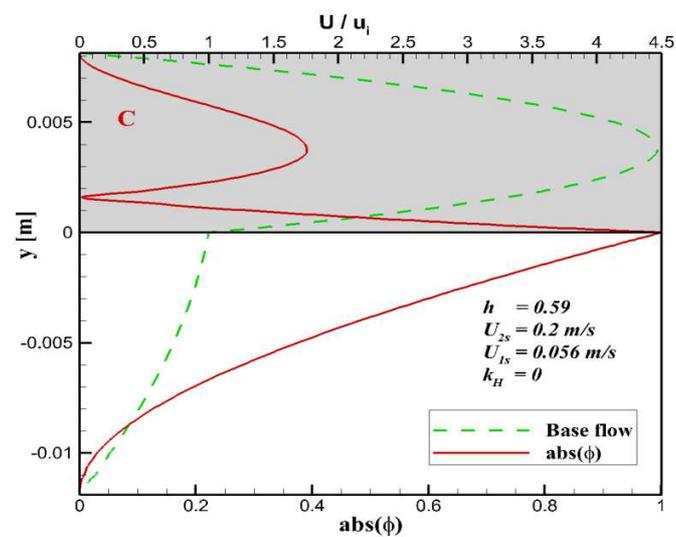
(e)

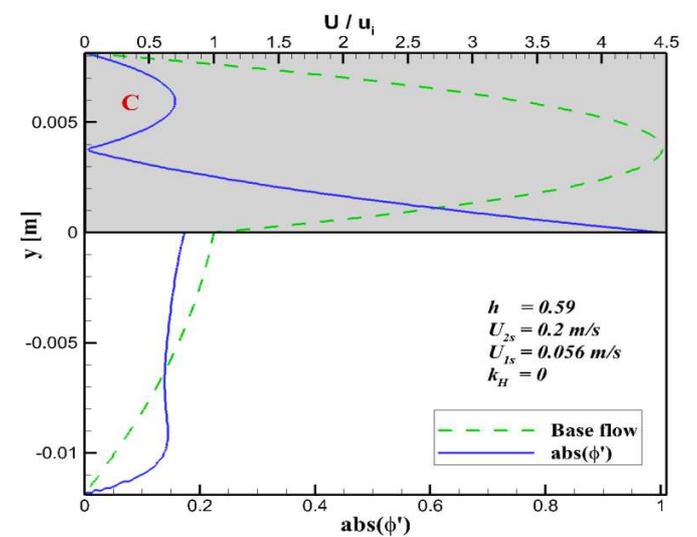
(f)

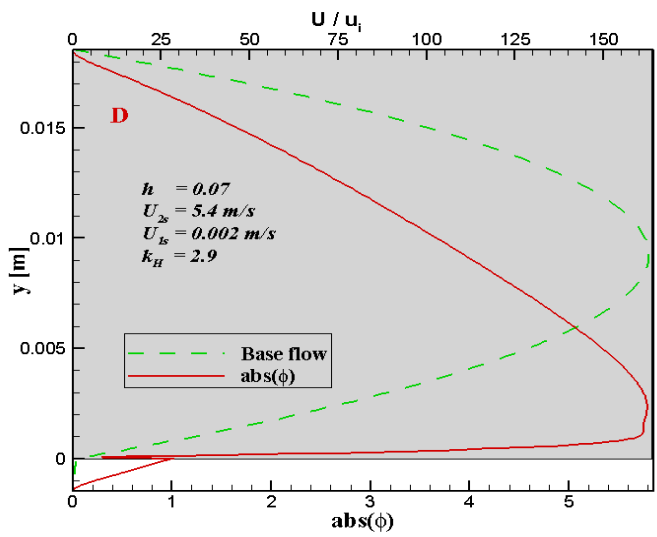

(g)

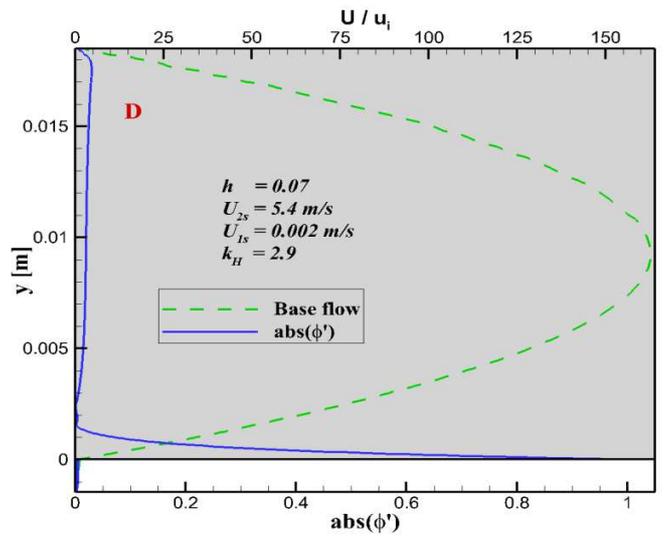

(h)

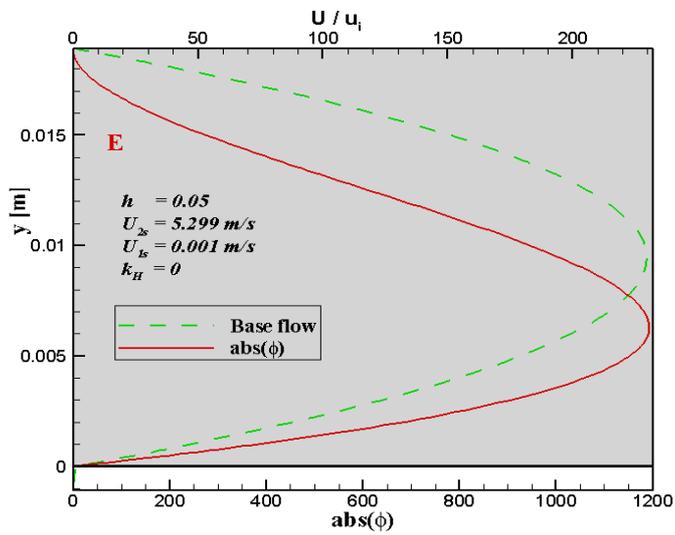

(i)

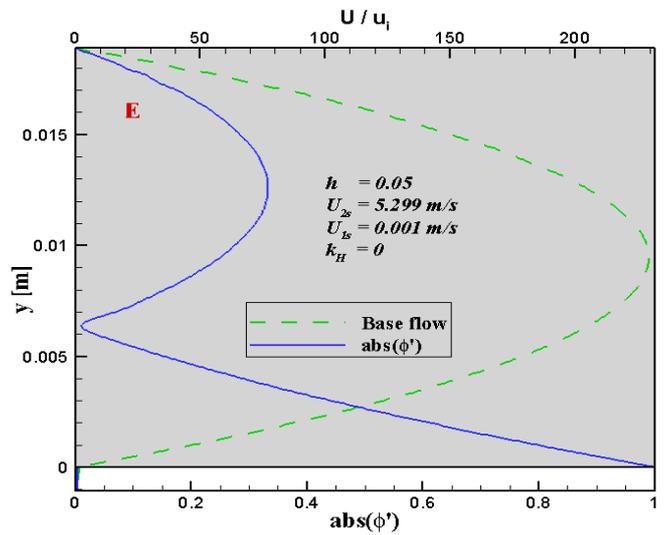

(j)

FIG. 13. Amplitudes of the critical perturbations of the stream function ((a), (c), (e), (g), (i); eigenfunction $abs(\phi)$, red solid lines) and its derivative ((b), (d), (f), (h), (j); $abs(\phi')$, blue solid lines) and base flow velocity profile ($U/u_i$, green dashed lines) at points A-E (see Figure 12); y < 0 – water; y > 0 (shaded region) – air.

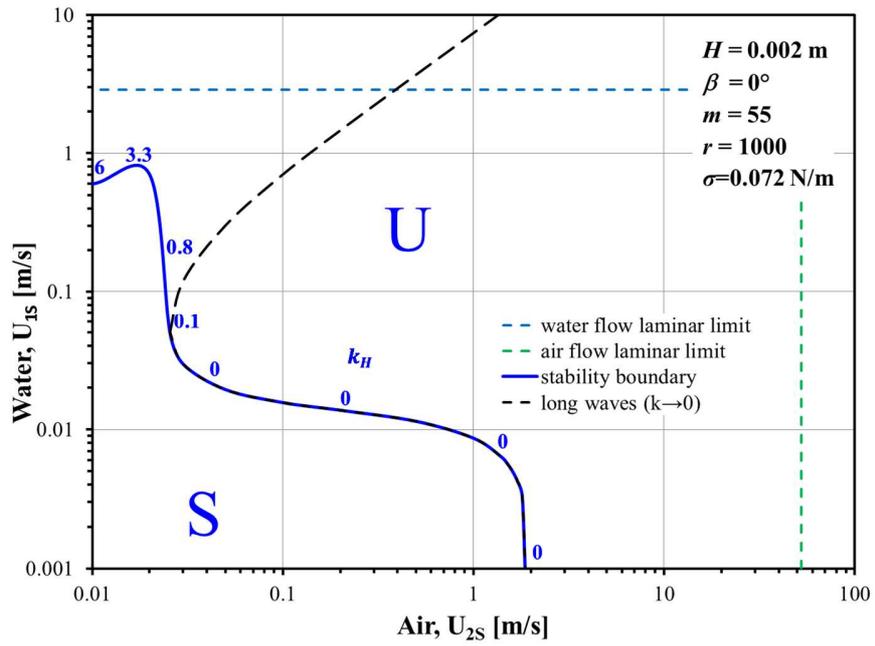

(a)

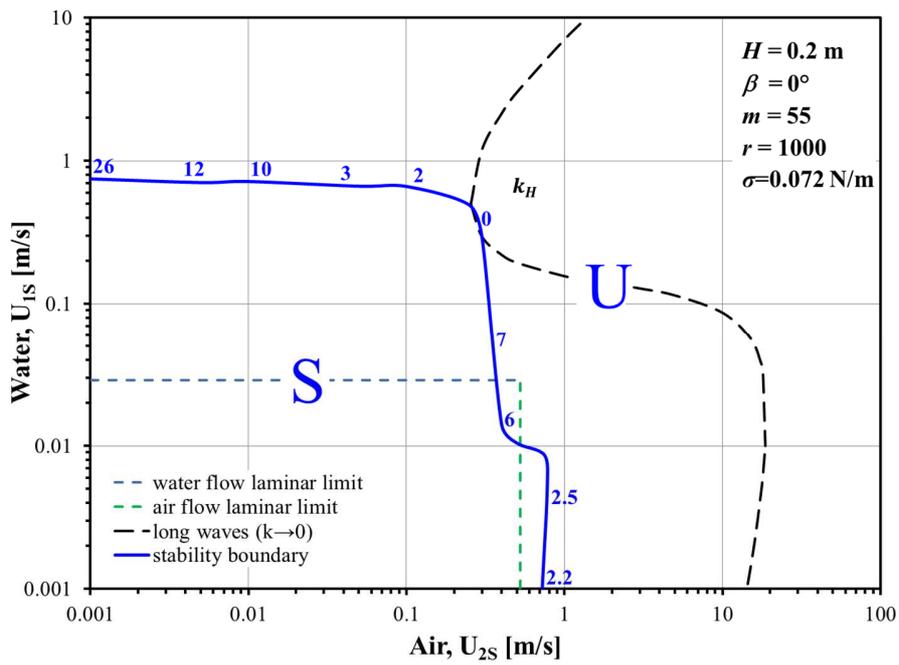

(b)

FIG. 14. Stability maps for air-water flow in a horizontal channel: (a) of 2 mm height; (b) of 200 mm height.

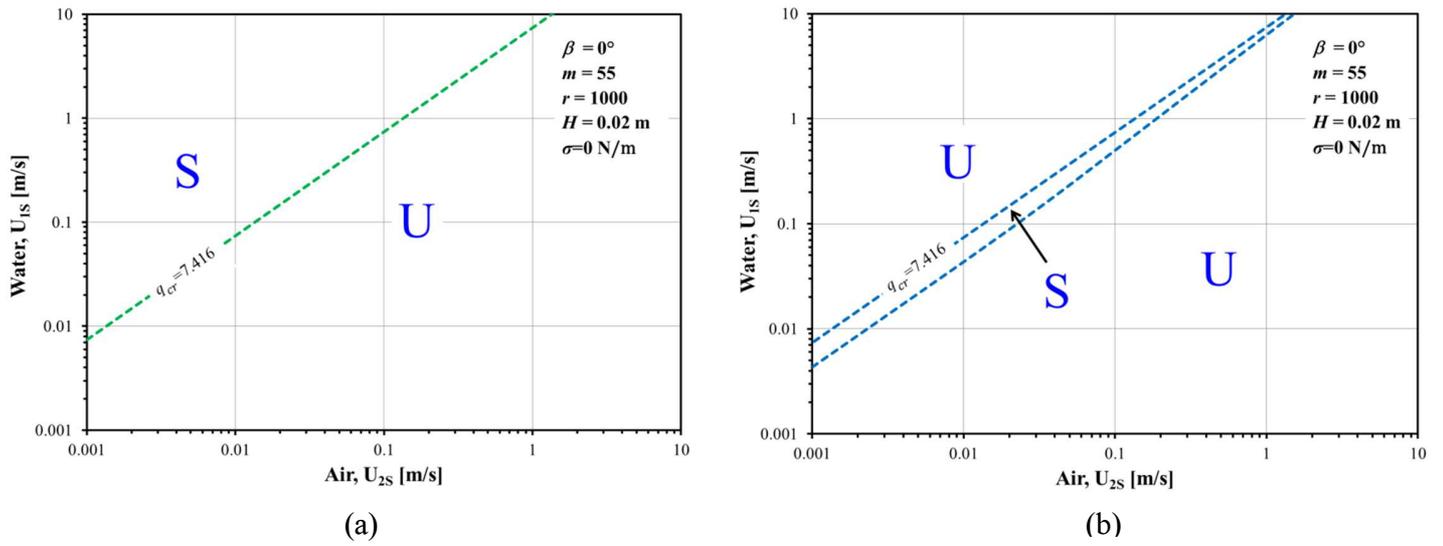

(a)                          (b)

FIG. 15. Stable regions for (a) infinite long waves $(k \to 0)$ and (b) short waves $(k = 10)$. Air-water flow without surface tension under zero-gravity condition is unstable for all flow rates for very short waves $(k \to \infty)$, hence, for all wavenumber perturbations.

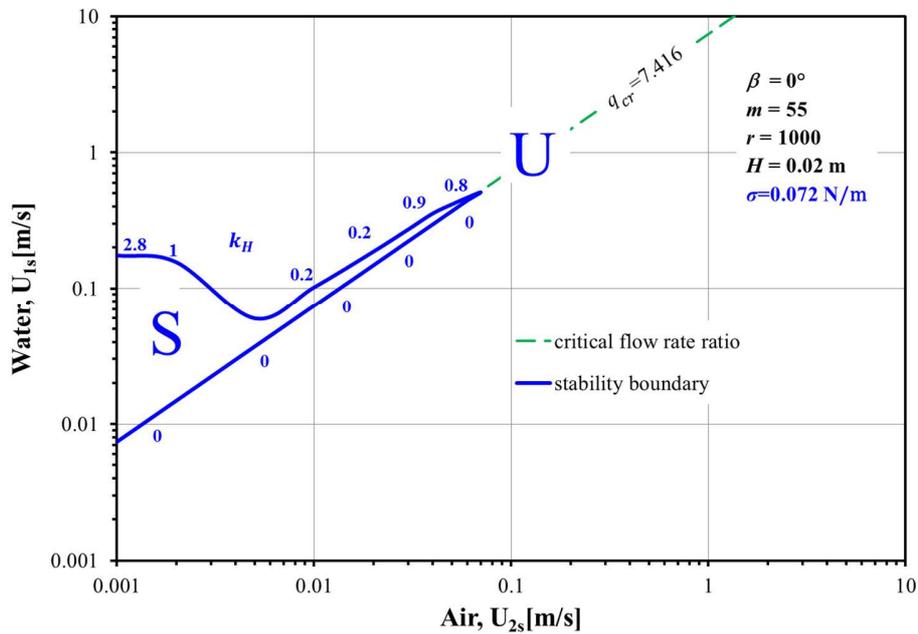

FIG. 16. Stability map for air-water system with surface tension under zero-gravity condition.

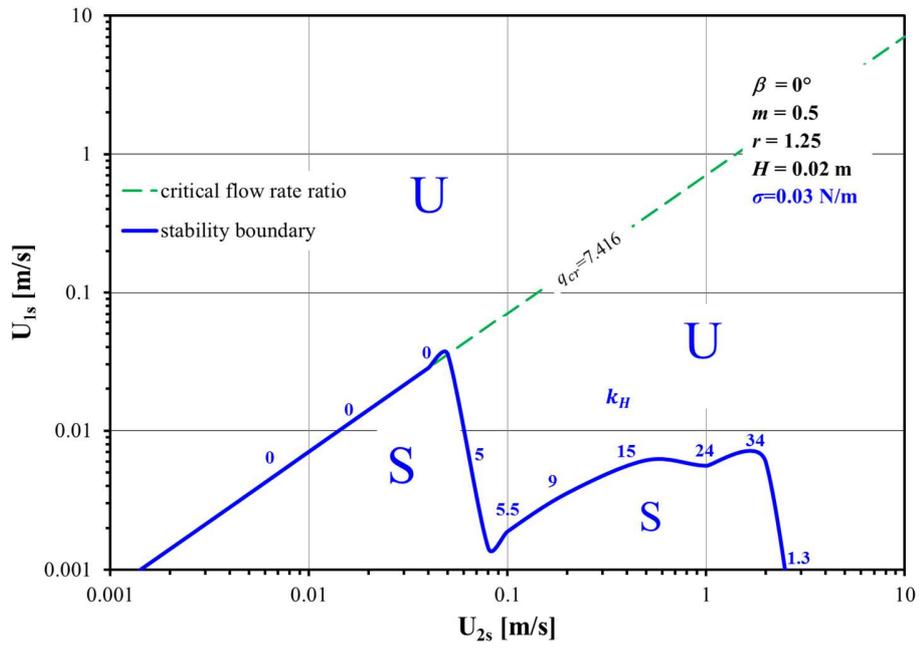

FIG. 17. Stability map for liquid-liquid system with surface tension under zero-gravity condition $(m = 0.5)$.